\documentclass[aps, prx, twocolumn, showpacs, superscriptaddress, floatfix, amsmath, amssymb]{revtex4-2}
\pdfoutput=1
\usepackage[utf8]{inputenc}
\usepackage{amstext}

\usepackage{hyperref}

\usepackage{graphicx}
\usepackage{xcolor}

\usepackage[english]{babel}
\usepackage{microtype}

\usepackage{esint}
\usepackage{bm}

\usepackage{hyperref} 
\hypersetup{
	pdftoolbar=true,        		 
	pdfmenubar=true,     		 
	pdffitwindow=false,     	 
	pdfstartview={FitH}, 
	pdftitle={},  
	pdfauthor={},     		 
	pdfsubject={},   			 
	pdfcreator={},   	        	   	
	pdfproducer={}, 		
	pdfnewwindow=true,      	
	colorlinks=true,       		
	linkcolor=black,          	
	citecolor=blue,        		
	filecolor=magenta,    		
	urlcolor=blue          		
}

\usepackage{stmaryrd}
\newcommand{\CCQ}{Center for Computational Quantum Physics, Flatiron Institute, 162 5th Avenue, New York, NY 10010, USA}

\newcommand{\Grenoble}{Univ. Grenoble Alpes, CEA, Grenoble INP, IRIG, Pheliqs, F-38000 Grenoble, France}
\newcommand{\Neel}{Univ. Grenoble Alpes, CNRS, Institut N\'eel, 38000 Grenoble, France}

\begin{document}

\author{Matthieu Jeannin}
\affiliation{\Grenoble}
\affiliation{\CCQ}
\email{matt.jeannin@gmail.com}

\author{Yuriel N\'u\~{n}ez-Fern\'andez }
\affiliation{\Neel}

\author{Thomas Kloss}
\affiliation{\Neel}

\author{Olivier Parcollet}
\affiliation{\CCQ}
\affiliation{Universit\'e Paris-Saclay, CNRS, CEA, Institut de physique th\'eorique, 91191,
	Gif-sur-Yvette, France}

\author{Xavier Waintal}
\email{xavier.waintal@cea.fr}
\affiliation{\Grenoble}

\date{\today}	
\begin{abstract}

We present a comprehensive set of numerically exact results for the Anderson
model of a quantum dot coupled to two electrodes in non-equilibrium regime.  We
use a high order perturbative expansion in power of the interaction $U$,
coupled to a cross-extrapolation method to long time and large interaction.
The perturbative series is computed up to $20-25$ orders, using tensor cross-interpolation. 
We calculate the full Coulomb diamond bias voltage - gate voltage map,
including its Kondo ridge, that forms the standard experimental signature of
the Coulomb blockage and the Kondo effect.  We present current-voltage
characteristics that spans three orders of magnitude in bias voltage and
display five different regimes of interest from probing the Kondo resonance at
small bias to saturation at very high bias.  Our technique also naturally
produces time-resolved interaction quenches which we use to study the dynamics
of the formation of the Kondo cloud. Finally, we predict several qualitatively
new physical features that should be within reach of existing or upcoming
experiments.

\end{abstract}
	
\title{A comprehensive study of out-of-equilibrium Kondo effect and Coulomb blockade }
\maketitle
%\author{Matthieu JEANNIN}

\section{Introduction}

\begin{figure} \centerline{	\includegraphics[scale=0.55]{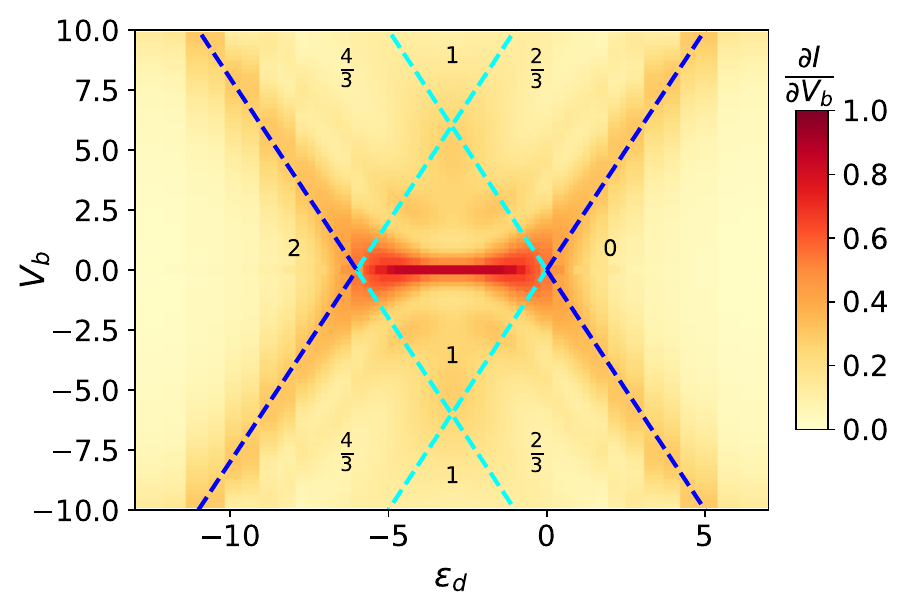}
      } \caption{ 
	 Differential conductance through the quantum dot computed using the TTD method for the Anderson model, at zero temperature, for $U =6$ 
	 as a function of the gate voltage $\epsilon_d$ and bias voltage $V_b$. 
	 Dashed lines indicate the Coulomb diamonds (as given by the semi-classical, sequential
	 tunneling, Cf Appendix \ref{sec:semi-classical}). 
	 Blue (resp. cyan) lines mark its outer (resp. inner) edges.
	 Numbers are the charge occupation predicted by the sequential model.
	 The Kondo ridge is clearly visible at $V_b = 0$ in the one electron regime.
     } \label{fig:diff_Kond_6}
\end{figure}

Some figures display features that are so robust and so proeminent that they eventually get a name.
Quantum dots display \emph{two} manifestations of many-body effects, due to the Coulomb
repulsion, the experimental signatures of which received such labels.  First, the Coulomb blockade, as current only flows thought the
dot when the gate voltage is adjusted to allow charge fluctuations on the dot;
second, the Kondo effect, a true quantum many-body effect at low voltage and
energy, in which the emergence of the Kondo resonance at low temperature opens
a many-body conduction channel through the dot. These effects are well
understood and documented in experiments. 
They are clearly visible in the
differential conductance through the dot,  manifesting as the so-called "Coulomb diamond"
and "Kondo ridge" feature, as illustrated on Figure \ref{fig:diff_Kond_6}
(a third celebrated curve is the "Coulomb staircase" shown in Fig.\ref{fig:coulomb_blockade}).  The
Coulomb diamonds has been observed in countless samples including GaAs/AlGaAs
heterostructures \cite{Mak13}, silicon heterostructures
\cite{Hofheinz06,Klein07,Li13}, graphene \cite{Stampfer08,Guttinger09} and
other \cite{Dekker99,Park02,Leturcq09,Mol15,Kotekar-Patil19,Shibata23}. The
Kondo ridge has also been observed repeatedly, but it requires a fine control
of the coupling of the quantum dot to the electrodes. Examples include
AlGaAs/GaAs heterostructure \cite{Schroer06}, carbon nanotubes
\cite{JarilloHerrero05,Delattre09,Niklas16}, molecular quantum dots
\cite{Roch09,Guo21} and other systems
\cite{Hamaya07,Klein07,Kretinin11,Svilans18,Kurzmann21,Mittag21}.

In spite of numerous studies of the Kondo effect, 
a controlled computation of the differential conductance for a wide range of $\epsilon_d$ 
remains a significant challenge.
Indeed, it combines several difficulties:
{\it i)} the complexities of many-body physics—especially with
emerging low-energy scales involving long-distance phenomena;
{\it ii)} the difficulties inherent to out-of-equilibrium calculations;
{\it iii)} the speed and robustness requirements of the method in order to compute a large number of values of
the gate voltage $\epsilon_d$ and the bias voltage $V_b$.

Fortunately, in the last decade we have entered a new era with the emergence of 
{\it precision quantum many-body methods}, which aim to provide controlled and predictive
computations of the properties of quantum many-body systems, beyond qualitative pictures.
The goal of this article is to leverage on such a technique, 
the tensor train diagrammatics (TTD) recently introduced by some of us \cite{NunezFernandez22}, 
to provide a comprehensive calculation of the
differential conductance map in quantum dots, which is presented on Figure \ref{fig:diff_Kond_6}.

The minimal model for the quantum dot is the single impurity Anderson model \cite{Anderson61}
(\textit{SIAM}). It consists in one interacting site (with the Coulomb repulsion $U$)
coupled to a non-interacting fermionic bath.  Originally introduced to model impurities in
metals, the SIAM has become central to mesoscopic physics.
%, particularly in the study of quantum dots
\cite{Meir93,Beenakker91,Cronenwett98,Goldhaber-Gordon98}. It is also a
building block for quantum embedding methods \cite{Georges96, kotliarRMP2007,
Aoki14} and is relevant to experiments in both steady-state mesoscopic systems
\cite{Cronenwett98,Goldhaber-Gordon98} and transient out-of-equilibrium systems
such as ultracold atoms \cite{Kanasz-Nagy18,Schafer20}.

In equilibrium, both numerical and analytical methods deliver quantitative
results for the SIAM
\cite{Wilson75,Krishna-murthy80,Tsvelick83,Hirsch86,Bulla08,Grundner23},
including the commonly used Quantum Monte Carlo algorithm in continuous time
\cite{Rubtsov05, Werner06, Gull08, Gull11}.

In out-of-equilibrium regimes however, quantitative results are more difficult
to obtain. Many approximate approaches have been proposed to solve the
out-of-equilibrium SIAM model, often relying on uncontrolled approximations
\cite{Meir93,Wingreen94,Eckstein10,Spataru09,VanRoermund10,Nuss12,Souto18,Konik01,Konik02,Chao11,Anders05,Anders06,Anders08,Anders10,Schoeller00,Jakobs10,Hershfield91,Fujii03,Dorda14,Mora15}.
Numerically exact algorithms provide error bounds that can be made
arbitrarily small with increased computational resources, but remain limited by
factors such as the strength of interaction or short time scales, e.g.
Tensor Network methods \cite{Anders05,Dias08,Heidrich-Meisner09,Nuss15,Kohn22,Wauters24},  
Numerical Renormalization Group (NRG) and time-dependent
density matrix renormalization group (t-DRMG) methods \cite{Schwarz18,Lotem20},
%represent promising avenues of research. In the thermodynamic limit, exact
%methods typically rely on some form of expansion, with each order computed as a
%multidimensional integral using Monte Carlo methods. This includes 
path integral techniques \cite{Weiss08, Segal10, Thoenniss23, Chen24},
continuous-time Monte Carlo methods for both imaginary \cite{Han07} and
real-time \cite{Werner10} calculations, and Inchworm algorithms
\cite{Cohen14,Cohen15}.

Among the numerically exact approaches for non-equilibrium systems, we focus on
a systematic high-order perturbation expansion in the coupling constant $U$,
coupled with efficient resummation techniques. This method, initially developed
\cite{Profumo15,Bertrand19,Bertrand20,Macek20} as a real-time, non-equilibrium
extension of diagrammatic QMC \cite{Prokofev98,VanHoucke10}, has proven capable
of computing both transient and long-time steady-state regimes
\cite{Profumo15,Bertrand19,Bertrand20}. Despite its apparent conceptual simplicity, it has
shown remarkable success in strongly correlated regimes, such as in describing
the Kondo effect.

A major recent progress for this method has been the replacement of Monte-Carlo
by a tensor network technique, i.e. the tensor cross-interpolation (TCI), to
compute the high dimensional integrals expressing the perturbative coefficients
\cite{NunezFernandez22}.  Not only is this Tensor Train Diagrammatic (TTD)
method much faster and precise than Monte-Carlo, it is also immune to the ``sign problem".
Hence, it enables us to obtain many more orders of the perturbative expansion
of physical observables (up to order $30$) over a wide range of regimes.

In parallel, many improvements have been proposed in resuming perturbation
series in order to calculate physical observable from their expansion
coefficients.  Previous works used Lindelöf and Padé resummation
\cite{VanHoucke2012,Profumo15}, or methods based on the analytical structure of
the series \cite{Bertrand19}.  Recently, exploiting a remarkable low-rank structure
of physical observables as functions of interaction energy and time $(U,t)$,
we introduced another approach for the resummation of the perturbative
series: the cross-extrapolation \cite{Jeannin24}. It makes
extrapolations of perturbative series to large couplings and times possible,
more automatic and more robust.

In this paper, we use the TTD  and the cross-extrapolation methods to obtain a
comprehensive set of results on the quantum dot, modeled by the SIAM.  We
compute the charge on the dot $Q$, the current through the dot $I$ and the
magnetization $M$ in magnetic field, as a function of the bias voltage $V_b$,
the gate voltage $\epsilon_d$ and the interaction $U$.  We present a detailed
map of the Coulomb blockade, and the Kondo ridge at low voltage and
temperature.  We also study in detail the relaxation of the transient to the
non-equilibrium steady state.

This paper is organised as follows: in section \ref{sec:model}, the SIAM model
is presented, along with the most significant results from this article. In
section \ref{sec:I_Vb} we study the current/voltage characteristics of the
system. The transient behavior of the observables after a quench is given in
section \ref{sec:cur}. In section \ref{sec:Kondo} we discuss the various manifestation of the Kondo temperature. Finally, in \ref{sec:coulomb-blockade} we study the physics of the Coulomb blockade.

\begin{figure}
	\centering
	\includegraphics[width=0.45\textwidth]{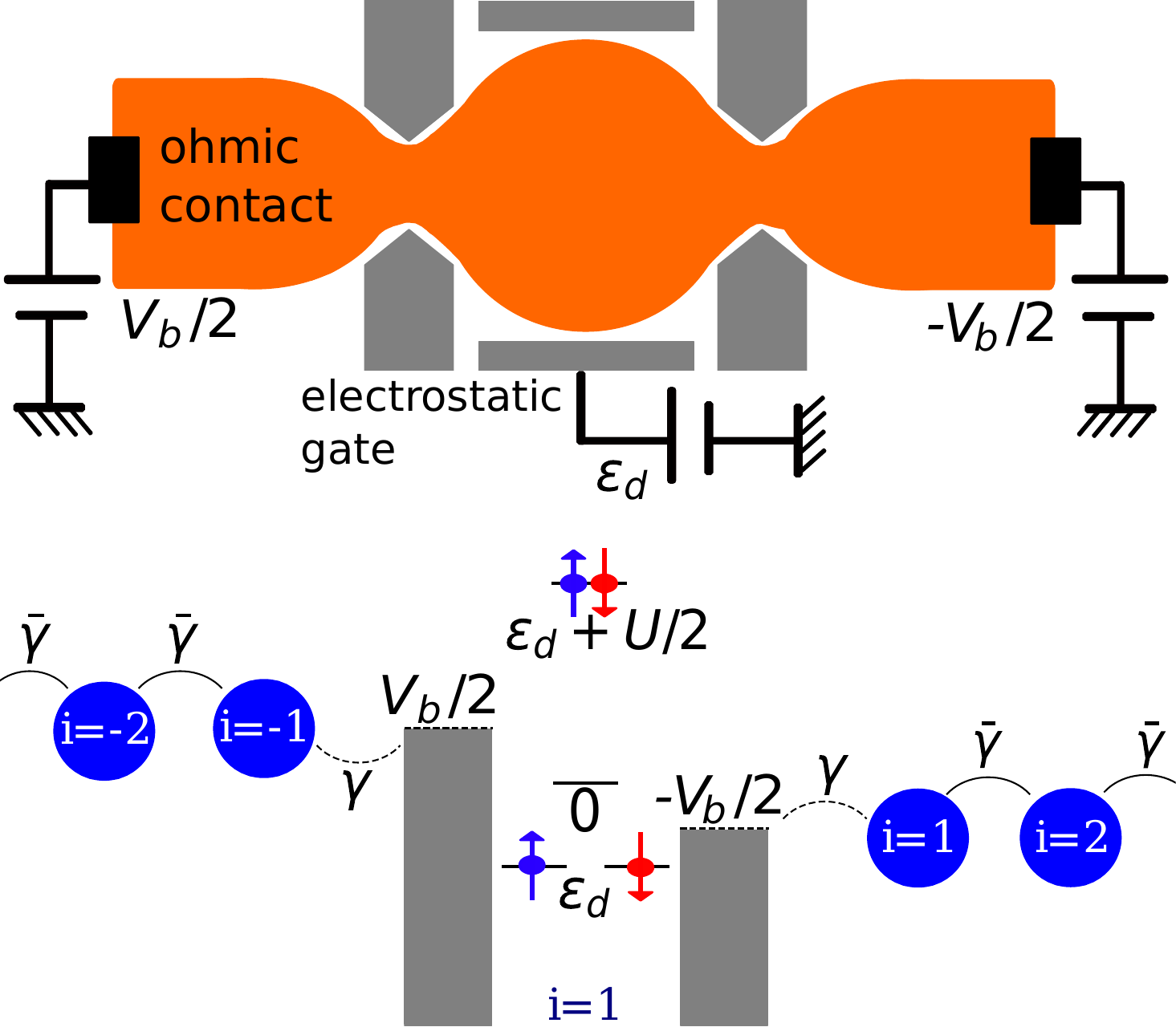}
\caption{Top: electrical diagram of the quantum dot coupled to two electrodes.
Bottom: schematic of the SIAM}
\label{fig:siam}
\end{figure}

%\section{From Coulomb blockade to Kondo physics.}
\section{Approach and results: an overview}
\label{sec:model}
Let us start with an overview of the model, the method and the main result
in the non-equilibrium steady state.

\subsection{SIAM in the stationary regime}

The quantum dot is modeled by a single interacting level 
(site $i=0$) weakly connected to two semi-infinite one dimensional electrodes ($i<0$ and $i>0$,
respectively) at chemical potential $\mu_{l/r} = \pm V_b/2$, as
depicted in Fig.~\ref{fig:siam}. 
Its Hamiltonian reads

\begin{equation}
\hat{H} =  \epsilon_d(\hat{n}_\uparrow+\hat{n}_\downarrow) + U \hat{n}_\uparrow\hat{n}_\downarrow
+ \sum_{i\sigma} \left(	\gamma_i c^\dagger_{i,\sigma 
	}c_{i+1,\sigma } +\text{h.c.}\right)
\end{equation}
where $c^\dagger_{i,\sigma}$ ($c_{i,\sigma}$) creates (destroys) an electron on site $i$ with spin $\sigma = \uparrow, \downarrow$,
$\epsilon_d$ is the on-site energy on the dot controlled by an electrostatic gate and $\hat{n}_{\sigma}= \hat{c}^\dagger_{0,\sigma}\hat{c}_{0,\sigma}$. 
We work directly in the thermodynamic limit with an infinite number of bath sites. 
The hopping between two neighbouring sites is denoted by $\gamma_i=\bar{\gamma}$,
except for the coupling to the dot $\gamma_0=\gamma_{-1}=\gamma$. 
Energies are expressed in unit of the tunneling rates to the dot $\Gamma = 2\gamma^2/\bar{\gamma}$. % times in unit of $\frac{\bar h}{\Gamma}$.
All calculations are performed at zero temperature. 
The charge on the dot and the current flowing through it are given by (Cf appendix \ref{sec:Lesser Green function and observables}): 
\begin{align}
	Q &= \sum_{\sigma=\uparrow,\downarrow}\langle \hat c_{0,\sigma}^\dagger \hat c_{0,\sigma} \rangle \\	
	I_r &=  -i \gamma  \sum_{\sigma=\uparrow,\downarrow} \left[ \langle \hat c_{1,\sigma}^\dagger \hat c_{0,\sigma} \rangle - \langle \hat c_{0,\sigma}^\dagger \hat c_{1,\sigma} \rangle\right] \label{eq:current_def}
\end{align}
Note that Eq. \eqref{eq:current_def} defines the current flowing from the dot to the
right electrode. A similar current $I_l$ can similarly be defined with the left electrode.
In steady state $I_r = - I_l = I$.
The conductance $\frac{I}{V_b}$ and
the differential conductance $\frac{\partial I}{\partial V_b}$ are expressed in units
of the conductance quantum 
$g_0\equiv \frac{2e^2}{h}$.
In the following, currents are expressed in units of $\frac{\Gamma g_0}{e}$.

% ---------------------------------------------------------------

\subsection{Summary of the Tensor train diagrammatics approach}

\begin{figure*}
	\centerline{
		\includegraphics[scale=0.9]{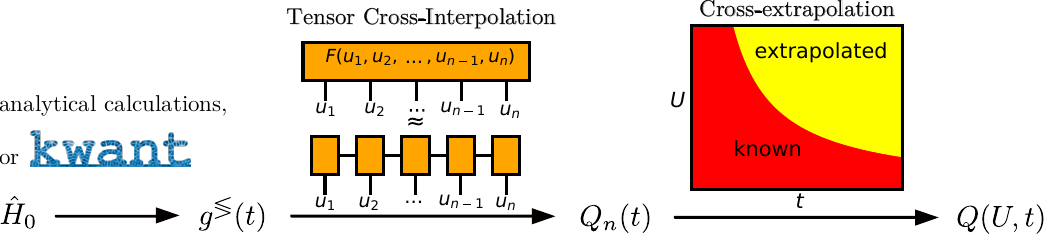}
	}
	\caption{ Overview of the TTD method. 
Starting from non interacting Green's functions, a decomposition of
the multidimensional integral is obtained using Tensor Cross-Interpolation
to compute the perturbative expansion, from which the physical observables
$Q(U,t)$ are reconstructed as a function of interaction $U$ and time $t$ using
cross-extrapolation.}
	\label{fig:method}
\end{figure*}

We study the out-of-equilibrium SIAM with the tensor train diagrammatic (TTD) approach introduced in \cite{NunezFernandez22, Jeannin24} and depicted in  Fig.\ref{fig:method}.
The goal is to calculate observables, such as the current $I(t,U)$ or the charge $Q(t,U)$, after an interaction quench at $t=0$ in terms of the series
\begin{equation}
Q(U,t) = \sum_{n=0}^{+\infty} Q_n(t) U^n.
\label{eq:series}
\end{equation}
the method consists of three main steps: {(\it i)} solve the non-interacting model
to obtain the corresponding non-interacting Green's functions (such as $g^<(t)$).
{(\it ii)} Calculates the coefficients $Q_n(t)$ with high precision 
and {(\it iii)} resumming this perturbative expansion.

The first step is carried out either analytically in simple models (Cf Appendix \ref{sec:Non-interacting Green})
or numerically for more complex geometries using e.g. the Kwant \cite{Groth14} or Tkwant \cite{Kloss21} software.
Note that the non-interacting Green's function may already be out-of-equilibrium in presence of a bias voltage $V_b$.

Second, we perform the computation of the 
perturbative expansion within the Keldysh formalism 
in powers of $U$. 
The order $n$ term $U^n Q_n(t)$ is given by an $n$-dimensional integral
that contains $2^n$ terms constructed out of products of non-interacting
Green's functions \cite{Profumo15,Bertrand20}. We compute this integral using
the Tensor Cross-Interpolation algorithm which vastly outperforms previous
quantum Monte-Carlo and quasi Monte-Carlo approaches for this problem
\cite{NunezFernandez22,NunezFernandez23}, see \cite{NunezFernandez24}  for an
open source library.
We calculate all orders up to $n=N$ with typically $N=20-25$.

The last step of the method consisting in summing the perturbative series
\begin{equation}
Q(U,t) = \sum_{n=0}^{+\infty} Q_n(t) U^n \approx \sum_{n=0}^N Q_n(t) U^n.
\label{eq:series_approx}
\end{equation}
at any time $t$ and at large interaction $U$.
As we switch on the interaction at $t=0$, 
we obtain both the transient and the steady state  $t\rightarrow \infty$ ($t\gg N/\Gamma$) 
regime (the long time limit can be taken order by order in perturbation in $U$).
The behaviour of the series is quite different in these two regimes.
At any fine time $t$, it is absolutely convergent since $Q_n \sim t^n/n!$ \cite{Bertrand19}. 
However, at least $N \sim O(Ut)$ terms are required to converge the sum. 
On the other hand, in the steady state, it has a finite radius of convergence $R$ and $Q_n\sim 1/R^n$.
Hence a naive summation fails for $U>R$.
Conformal transformations \cite{Bertrand19} can overcome this difficulty and resum the series in 
interesting strongly correlated regimes. However its success strongly depend on the analytical structure of $Q(U,t=\infty)$ and 
it fails in some important regimes \cite{Jeannin24}.

In this work, we follow the {\it cross-extrapolation} approach \cite{Jeannin24}.
We obtain $Q(U,t)$ at large time and small interaction $U<R$ or small time and large interaction $U\sim N/t$ using a plain summation of the series. 
Then we simultaneously extrapolate the small $U$ data to larger values
\emph{and} the small values of $t$ to larger values with the constraint that the
final result at a given $(U,t)$ should be the consistent.
The success of the approach relies on an underlying structure, the low rank structure of $Q(U,t)$.
We observe that this procedure 
is more systematic and  robust in practice than the conformal mapping of \cite{Bertrand19}, 
even though we will exhibit later one case where the low-rank assumption fails.

\subsection{Main results in the stationary limit.}

The main effect of interaction in a quantum dot is the well known {\it Coulomb blockade} phenomena.
The current can only flow through the dot when the bias voltage $V_b$
is larger than the energy needed to add an electron in the dot (here from zero to one or one to two).
At small $V_b$, the current can flow only at $\epsilon_d=0$ and $\epsilon_d = -U$. 
The corresponding sequential tunneling \cite{Beenakker91} limit is described in Appendix \ref{sec:semi-classical}.
At larger $V_b$, the current flows in two regions known as the Coulomb diamonds, respectively for $2|\epsilon_d| \le|V_b|$ and $2|\epsilon_d+U| \le |V_b|$.
They have been observed in a very large number of experiments with quantum dots made
in materials as diverse as GaAs \cite{Mak13}, carbon nanotubes \cite{Dekker99}, semiconducting nanowires \cite{Hofheinz06} and more \cite{Stampfer08,Leturcq09,Kotekar-Patil19}.
The Coulomb blockade is well understood, in particular starting from an isolated dot limit.
It is a correlated effect in the sense that it is not described by a Hartree-Fock mean field calculation: 
the current is correlated to the (fully quantized in the sequential tunneling theory)
number of electrons in the dot. 

A more subtle manifestation of correlations is the {\it Kondo effect} \cite{Kondo64,Glazman88,Lee92,Goldhaber-Gordon98}, 
at low energy, for frequencies and temperatures lower than the Kondo temperature $T_K$.
The {\it Kondo ridge} is its main experimental manifestation:
at $V_b \ll T_K$, the differential conductance  $\partial I/\partial V_b$  is non zero in the entire region $-U\le \epsilon_d \le 0$.
Furthermore the transmission becomes perfect in the $V_b\rightarrow 0$ limit and the conductance equals to the quantum of conductance $g_0$ \cite{Cronenwett98}.

Figure \ref{fig:diff_Kond_6} shows our calculation of the differential
conductance as a function of the on-site energy $\epsilon_d$ and the voltage
bias $V_b$ for $U=6$.  The plot is made of $100\times 100$ computed points,
requiring calculations at order $N=17-23$ (more orders
are needed close to the Kondo ridge).  The Kondo ridge (the dark red region) is
clearly visible, with its widening around the degeneracy points. 
  As expected, it is only present when there is an odd (here one) number of
electrons in the dot (Kondo valley) and disappears for $\epsilon_d<-U$ (two
electrons) and $\epsilon_d>0$ (zero electrons).  Note the particle-hole
symmetry around $\epsilon_d=-U/2$. 

The Coulomb diamonds are also clearly visible underneath the predictions
 of the sequential tunneling theory (highlighted by the dashed blue lines, both dark and cyan).
According to the semi-classical sequential tunnelling theory
\cite{Beenakker91} (Cf appendix \ref{sec:semi-classical}), sharp
variations in the differential conductance are expected when the electrode
energies align with the energy levels of the quantum dot (indicated by dashed
lines of both colors on Fig. \ref{fig:diff_Kond_6} ). 
Our numerically exact results present however a different picture.
While the outer resonance lines (dashed dark blue) are clearly observed,
the resonance inside the diamond-shaped region (dashed cyan) are barely visible.
This is in excellent agreement with experiments \cite{Shin09} and indicate that above and below the Kondo ridge, quantum fluctuations continue to play an important role in the physics.

Let us emphasize that, while a small fraction of these results has or could have been obtained previously
with diagrammatic Monte-Carlo techniques \cite{Bertrand19, Macek20}, 
only the TTD approach allows to compute in all parameters regimes, 
including those in which the $N$ dimensional integrals of the perturbative expansion are very
oscillatory. Indeed, the TTD is insensitive to the ``sign problem" \cite{NunezFernandez22}.

\begin{figure} \centerline{
   \includegraphics[scale=0.6]{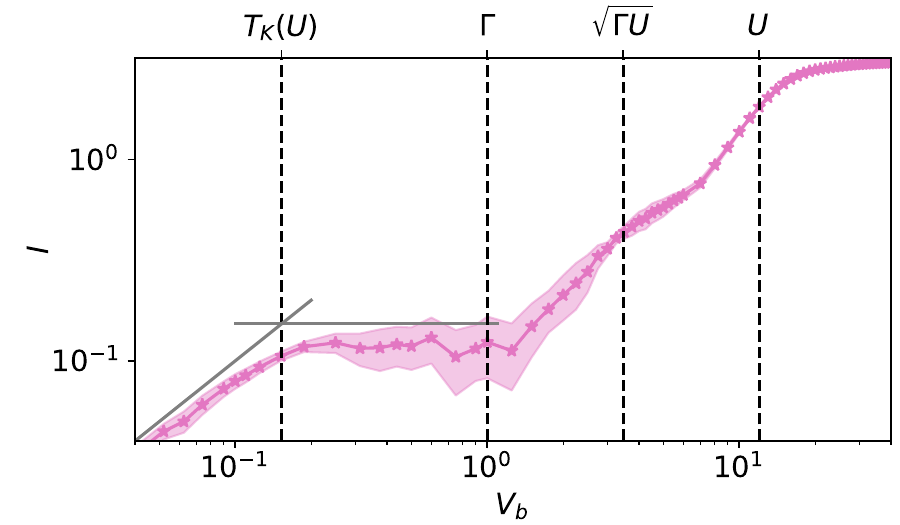}} 
   \caption{ Current $I$ as a function of $V_b$ at $U=12$. The dashed lines
   indicate the position of the energy scales discussed in the text.  The grey
lines correspond to $I=V_b$ and $I=T_K(U)$. $N=23$ coefficients were used. The
Kondo temperature is extracted from Bethe Ansatz, as in \cite{Bertrand19}.
Error bars are obtained via cross-extrapolation \cite{Jeannin24}.
}
\label{fig:Kondo_I_Vb_regimes}
\end{figure}  

The SIAM problem exhibits \emph{four} different energy scales with a very different dependence on the interaction $U$ at large $U$:
the (exponentially) small Kondo temperature $T_K$, the tunneling rate to the electrode $\Gamma$, 
the energy associated with the fluctuations of the charge in the dot $\sqrt{\Gamma U}$ (as predicted by Bethe Ansatz \cite{Tsvelick83})
and finally the largest energy scale $U$ itself. 
At $U=12$ (the largest value for which we can perform the cross-extrapolation), 
these scales delimits well separated regimes for the current $I(V_b)$.
Our TTD method is able to resolve these different regimes  
as illustrated on Fig. \ref{fig:Kondo_I_Vb_regimes},
which shows  $I(V_b)$  in the middle of the Kondo ridge at the particle-hole symmetry point $\epsilon_d = -\frac{U}{2}$.
At low voltage, we observe perfect transmission $I=2e^2V_b/h$ for $V_b<T_K$ (black line) after which
$I$ reaches a plateau for $T_K< V_b <\Gamma$.
It increases again for $V_b >\Gamma$, seems to present a inflexion point close to $V_b\sim \sqrt{\Gamma U}$, 
and finally saturates for $V_b>U$ at $I = \frac{e^2\Gamma}{2\hbar}$. 
Notice that Fig.~\ref{fig:Kondo_I_Vb_regimes} spans three orders of magnitude in bias voltage.

\begin{figure}
\centerline{		
\includegraphics[scale=0.42]{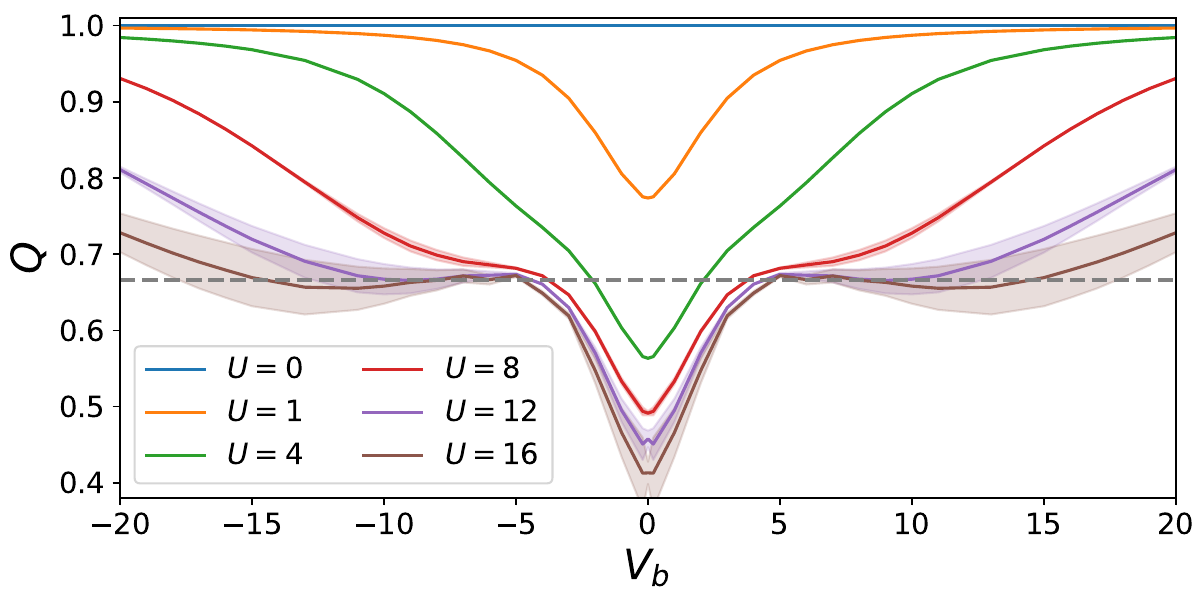}}
\caption{Charge blockade. Charge as a function of $V_b$ for different values of $U$ at $\epsilon_{d}=0$. The grey dashed line corresponds to the $Q=2/3$ plateau. $N=21$}
\label{fig:charge_Vb_ed0}
\end{figure} 

Finally, we present in Fig. \ref{fig:charge_Vb_ed0}
the dot occupation $Q(V_b)$ as a function of $V_b$ at $\epsilon_d=0$,  for different values of $U$ from $U=0$ to $U=16$.
At $U=0$, $Q=1$ for all $V_b$.
Some basic observations are in agreement with theoretical expectations 
given by the semi-classical approach of sequential tunnelling \cite{Beenakker91}, see appendix \ref{sec:semi-classical}.
In this limit, $Q=2/3$ independently of $V_b$ as soon as $V_b$ is non zero; 
while at large bias $V_b > 2U$, $Q=1$.
However $Q(V_b)$ forms a V-shape curve at large $U$ and small bias $V_b \le 4-5$.
This feature is a genuine
correlation effect that is \emph{not} present in  the standard sequential tunneling
theory.
The characteristic scale at which one observes the crossover from the V-shape
to the $2/3$ plateau is compatible with $\sqrt{\Gamma U}$ although more data
would be needed to confirm this point.

%%%%-===========================================================

\section{Current-Voltage characteristic}
\label{sec:I_Vb}
\begin{figure*}[h!]
\centerline{		\includegraphics[scale=0.48]{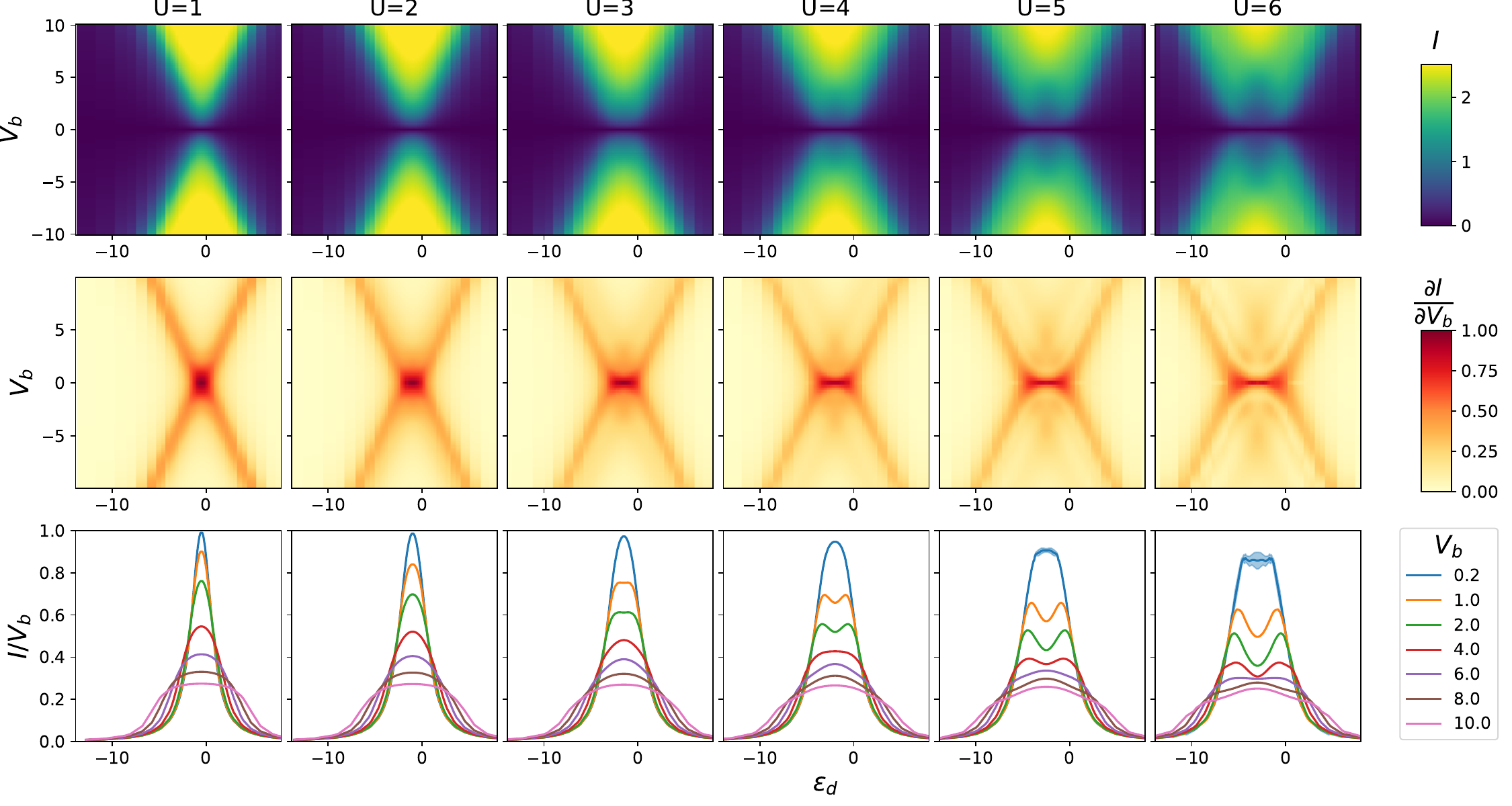}}
	\caption{Coulomb diamonds in the SIAM model in the steady state regime. $N=17$ coefficients are used, except for $V_b=0.2$ where  $N=23$. \textit{Upper panel - } color plot of the current as a function of $V_b$ and $\epsilon_{d}$ for different values of $U$. \textit{Middle line -} color plot of the differential conductance calculated by a polynomial fit to the current. \textit{Lower panel -} slices of the data showing the conductance $I /V_b$ as a function of $\epsilon_{d}$ for different values of $V_b$ and $U$.}
	\label{fig:coulomb_diamonds}
\end{figure*}  

In this section, we present additional results for the current voltage characteristics
in different regimes.  The current $I$ and differential conductance $dI/dV_b$
versus $\epsilon_d$ and $V_b$ are shown respectively in the first and
second line of Figure \ref{fig:coulomb_diamonds}. As one increases $U$, one
observes the separation of the two diamonds $2|\epsilon_d|<|V_b|$ and
$2|\epsilon_d+U|<|V_b|$, clearly visible in the current $I$ while the Kondo
ridge is clearly visible in the differential conductance. One also distinctly
sees that the Kondo ridge get thinner as one increases $U$. Its width is $\sim
T_K^{(1)}$ and will be studied quantitatively below.

The third line in Figure \ref{fig:coulomb_diamonds} shows the conductance $I/V_b$ versus $\epsilon_d$ for different bias values. The value of
the conductance should tend to $1$ (perfect transmission) for $V_b\ll T_K$ at 
the particle-hole symmetric point $\epsilon_d = - \frac{U}{2}$ \cite{Glazman88,Wingreen94}.
It is indeed the case for small $U$ (large $T_K$) but as one increases
$U$, the conductance gets smaller even for the smallest
value of $V_b$ shown in this set of data ($V_b=0.2$, additional data at smaller
bias will be shown below).  This is due to he fact that $T_K$ decreases and the condition
$V_b\ll T_K$ is no longer fulfilled.
At larger $V_b$ (for $V_b \gg T_K(U)$, the Kondo effect is suppressed, and two side peaks emerge, corresponding to Coulomb blockade.

\begin{figure*}
\centerline{	\includegraphics[scale=0.6]{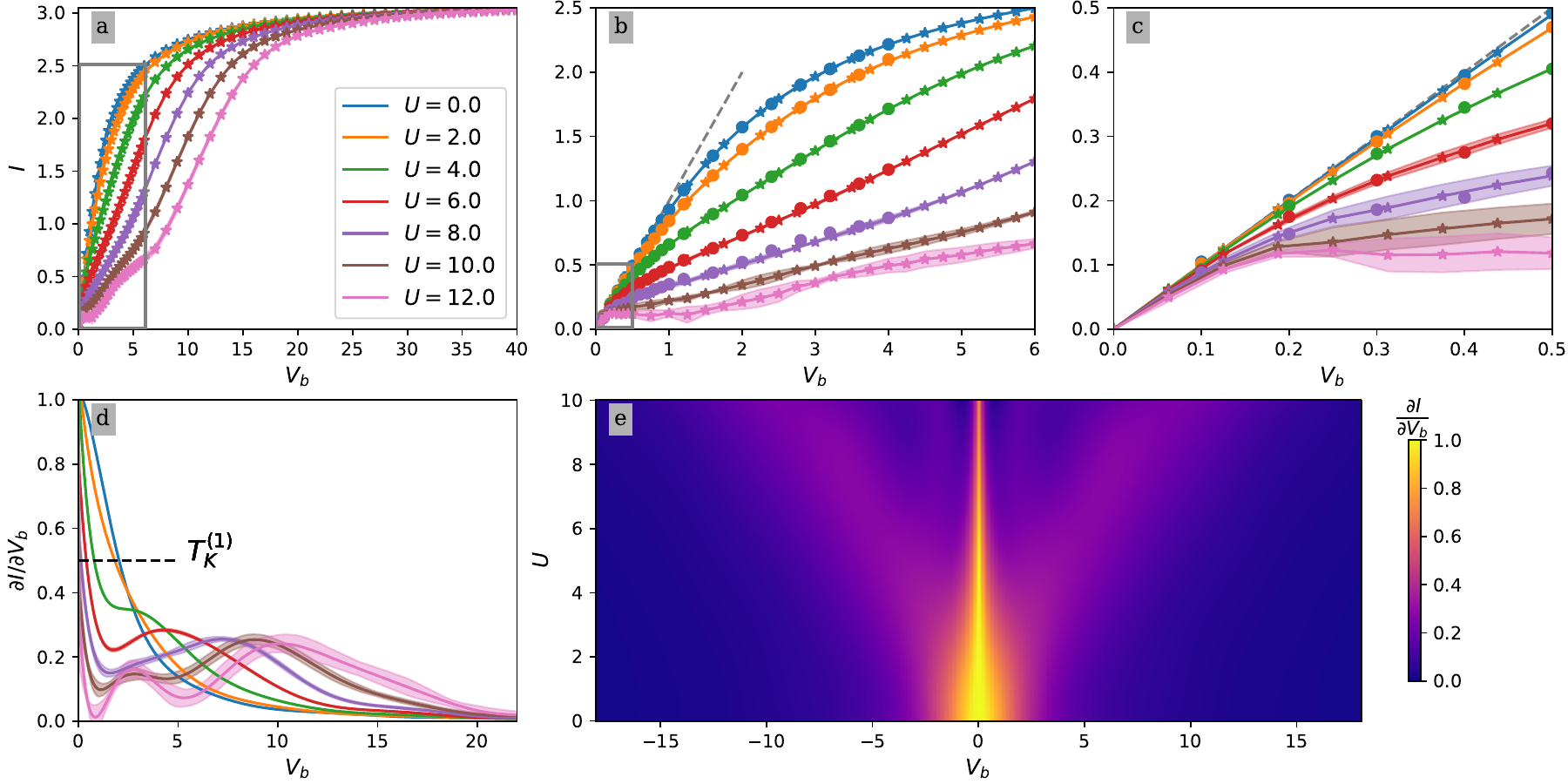}}
	\caption{ Current-voltage characteristic $I(V_b)$ at particle-hole symetric point $\epsilon_{d}=-U/2$ \textit{a,b,c)} Current as a function of the bias for different values of $U$. Stars corresponds to our results, and spots are benchmark from \cite{Bertrand19}. Panel (c) is a zoom of panel (b), panel (b) is a zoom of panel (a). \textit{d)} differential conductance as a function of the bais for different values of $U$. The differentiation is obtained by a polynomial fitting of the current, several fits have been performed to give error bars by bootstrapping. \textit{e)} Color plot of the differential conductance as a function of $U$ and $V_b$.}
	\label{fig:cur_I_Vb}
\end{figure*}    

In the particle hole-symmetry ($\epsilon_d = -\frac{U}{2}$) case,
the current-voltage characteristic $I(V_b)$ is presented on 
Figure \ref{fig:cur_I_Vb} for $U=0$ to $U=12$, 
with a zoom at low $V_b$ on different panels.
For $V_b <T_K(U)$, the current follows the $I=V_b$ line corresponding to perfect
transmission as a result of tunneling through the Kondo resonance. When $V_b >T_K(U)$ current gets out-of the Kondo regime and saturates at $V_b\sim U$.
The particle-hole symmetry point has been solved with diagrammatic QMC method \cite{Bertrand19}, 
and our TTD results (stars) are in excellent agreement with these previous results (circles).
This regime is quite favorable to QMC, as the perturbative integrals do not oscillate too much (absence of sign problem).
However, TTD still outperforms it (for a given precision
about twice as many orders can be calculated) and allows one to reach $U\sim 12$
when the previous calculation were limited to $U=8$. 

Figure \ref{fig:cur_I_Vb}d shows the
differential conductance as a function of the voltage bias, obtain
by differentiating a polynomial fit of  $I(V_b)$. Error bars are obtained by bootstrapping ($I(V_b)$ are generated by sampling the data within our error bars and used in the above mentioned procedure).
%(known forthe current); each of these curves are fitted, then differentiated; the
%sample-to-sample fluctuations of the obtain differential conductance is the
%error bar. 
Two peaks are observed: a first thin and intense peak around
$V_b=0$ reaching perfect conductance $\frac{2e^2}{h}$ corresponding to the Kondo
resonance, and a second broader peak for $V_b$ of the order of $U$ when the
singly occupied orbital levels and the Fermi level of the electrode are aligned
(classical lifting of the Coulomb blockade).  
% Cf relevant section
The width of the Kondo peak, defined as the value where the differential conductance reaches half of its maximum, provides an estimator 
of the Kondo temperature, denoted as $T_K^{(1)}(U)$.
%We define the Kondo temperature
%$T_k^{(1)}$ as the width of the Kondo peak, more precisely as the value of
%$V_b$ for which the differential conductance is $1/2$. 
%This is one of several
%compatible definitions of the Kondo temperature, see the discussion in section
%\ref{sec:Kondo}. 
Note that a third peak seems to be visible around $\sqrt{\Gamma
U}$ for the largest values  $U=10$ and $U=12$. Panel ({\sl e}) 
shows the differential conductance as a function of $U$ and $V_b$. The color plot clearly shows the
formation of the Kondo resonance as the central peak gets thinner for increasing
$U$. The secondary peaks around $|V_b|=U$ are also visible.

\begin{figure*}
\centerline{	\includegraphics[scale=0.6]{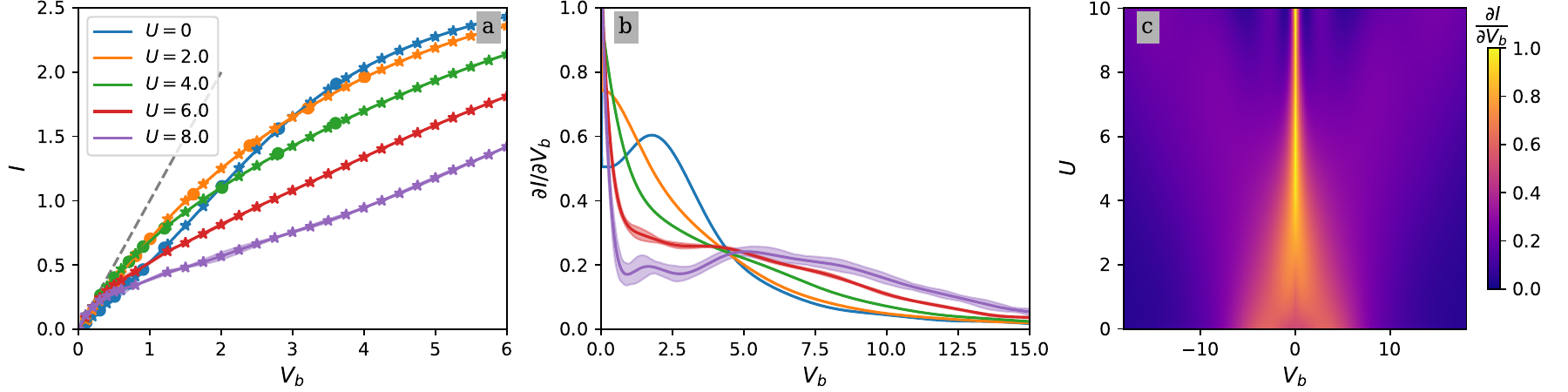}}
	\caption{ 
     Current-voltage characteristic $I(V_b)$ characteristic at $\epsilon_{d}=-U/2-1$ outside of the particle-hole symmetric point. Stars corresponds to our results, and spots are benchmark from \cite{Bertrand19}.  \textit{a)} Current as a function of the bias for different values of $U$. Spots are benchmark results from \cite{Bertrand19}. \textit{b)} differential conductance as a function of the bias for different values of $U$. The differentiation is obtained by a polynomial fitting of the current, several fits have been performed to give error bars by bootstrapping. \textit{c)} Color plot of the differential conductance as a function of $U$ and $V_b$.}
	\label{fig:cur_I_Vb_ed}
\end{figure*}

A computation away from half-filling is presented on 
Figure \ref{fig:cur_I_Vb_ed} for $\epsilon_d = -U/2 -1$.
As expected, for $U=0$ and $U=2$,  
the differential conductance (panel b) does no longer correspond to perfect transmission $V_b=0$
(the resonance is misaligned with the Fermi level, where the Kondo resonance forms).
A finite $U\approx 4$ is needed for the 
transmission to approach unity. % as is also visible on the color plot (panel c): the peak separates in two for small values of $U$. 

%%==================================================================================

\section{Transient current after a quench}
\label{sec:cur}

We now turn to the time-resolved dynamics of the system after a quench of the interaction
(together with the gate voltage and/or a Zeeman magnetic field). 
The TTD methods provides the full
time evolution of the observables in a single run, allowing to obtain the transient as well as the steady state \cite{NunezFernandez24}.

\subsection{Non Stationary SIAM}

We switch the interaction on at
$t=0$. The Hamiltonian $\hat H$ can be split into two terms:
\begin{equation}
\hat{H} = \hat{H}_0 + \hat{H}_{int}
\end{equation}
with 
\begin{align}
\hat{H}_0 &= E_d(\hat{n}_\uparrow+\hat{n}_\downarrow) + 
B_d(\hat{n}_\uparrow-\hat{n}_\downarrow) \\&+\sum_{i \in \mathbb{Z} \atop \sigma = \uparrow, \downarrow} 	\left(\gamma_i c^\dagger_{i,\sigma}c_{i+1,\sigma } + \text{h.c.}\right)
\end{align}
where $E_d$ is the initial energy of the dot and $B_d$ the initial (before the quench) Zeeman magnetic field and 
\begin{equation}
\hat{H}_{int}= U \theta(t) (\hat{n}_\uparrow-\alpha_\uparrow)(\hat{n}_\downarrow -\alpha_\downarrow ).
\end{equation}
With these definitions, the final gate voltage after the quench is 
\begin{equation}
\epsilon_d = E_d - \alpha U
\end{equation}
where
$\alpha = (\alpha_\uparrow+\alpha_\downarrow)/2$ while the final magnetic field is
\begin{equation}
b_d = B_d + U (\alpha_\uparrow-\alpha_\downarrow)/2
\end{equation}
For a given set of final parameters $(\epsilon_d,b_d)$, the shift $\alpha_\uparrow$ and $\alpha_\downarrow$ allow one to
perform different quenches from different initial states. This can be useful in
itself (if one is actually interested in the dynamics of different quenches) or because the
calculations are easier for some of these quenches (see \cite{Profumo15} and the discussion in Appendix \ref{sec:app:alphaRole}).

%% --------------------------------------

\subsection{Charge and current relaxation}
\label{sec:currentRelax}

\begin{figure*}
	\centerline{	\includegraphics[scale=0.51]{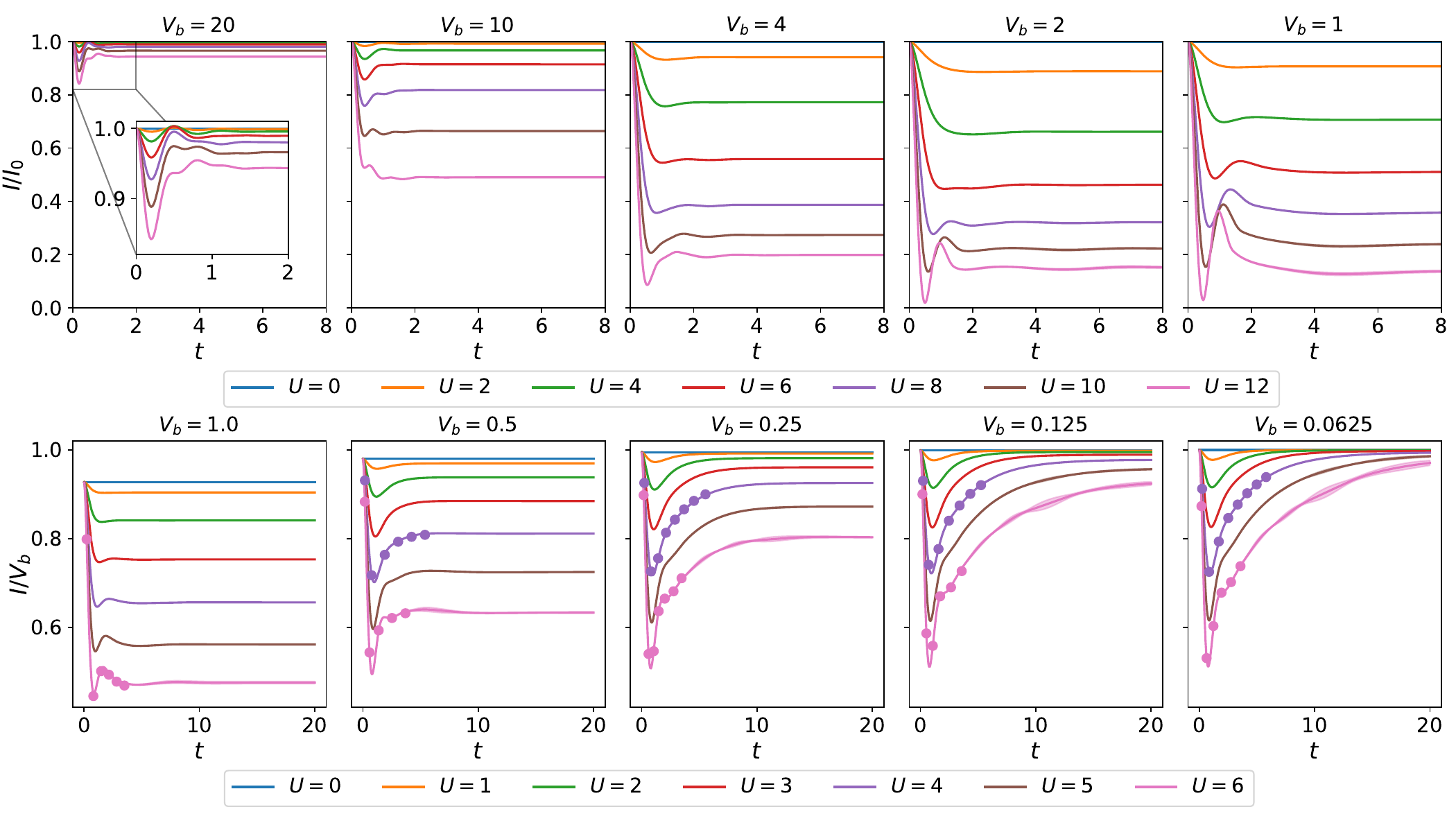}}
	\caption{Evolution of the current after a quench.
	The current is plotted as a function of time and $U$ for different values of $V_b$ using in the particle-hole symmetric regime $E_d=0$, $\alpha = 1/2$ and $N=23$. \textit{Upper panel} - current for high value of $V_b$, no Kondo effect is present and the current relaxes within $t\sim 4/\Gamma$, values up to $U=12$ can be obtained with reasonable error bars, current is normalized by the non-interacting current $I_0$. \textit{Lower panel} - Conductance for small values of $V_b$, the Kondo relaxation takes much more time (up to $t\sim 20/\Gamma$) and accessible $U$ values as much smaller, dots corresponds to values computed by Monte Carlo, see \cite{Werner10}.}
	\label{fig:diff_I_as_t}
\end{figure*}  

The time-resolved interaction quenches show a double time scales relaxation corresponding 
to the charge sector ($t\sim\hbar/\Gamma$) and spin sector ($t\sim\hbar/T_K$) that we now study.

Figure \ref{fig:diff_I_as_t} displays our data set for current versus time $t$
for different values of $V_b$ and $U$ at the particle-hole symmetric point
$\epsilon_{d} = -\frac{U}{2}$ evolving from a system without on-site energy
($E_d=0$, $\alpha=0.5$). The upper panels shows the larger values of $V_b$ (for
$t\le 8$) while the lower panels shows the smaller values $V_b\le 1$ (for
larger times $t\le 20$).

For the larger values of $V_b$, (upper panels of figure \ref{fig:diff_I_as_t}),
we plot $I(t)/I(t=0)$. The convergence with time is fast, i.e.
the stationary regime is already reached at $t\le 4/\Gamma$.
%after a
%reorganisation of the dot occupation (although $Q(U)=1$, for all $U$ at the
%particle-hole symmetry point).  
The calculations are also easier in that regime as the series converges quickly at small
time. Hence the upper panels includes values of $U$ up to $U=12$ while we
restricted the calculations to $U=6$ at very small biases (lower panels).
We also see that the interaction barely affect the current for the largest biases. 

At small bias $V_b \ll T_K$, the relaxation is much slower, it is limited by the time
it takes to form the Kondo cloud $\sim\hbar/T_K$.
After an initial charge equilibration for $t\sim 1/\Gamma$, 
we clearly see, in particular for large $U$, a long relaxation time
$t \sim \hbar/T_K^{-1}$ due to the formation of the Kondo resonance.
In the lower panel of Fig. \ref{fig:diff_I_as_t}, 
we plot $I/V_b$, which reaches unity (perfect transmission) in the Kondo regime. 
The capability to obtain the long time relaxation is one the main advantage of the TTD method:
for comparison, we also plot data points from \cite{Werner10}, obtained using Quantum Monte-Carlo
(full circles). We see that TTD can go to much longer time at small $V_b$, e.g. $\sim 20/\Gamma$ and relax to the stationary limit.

\begin{figure*} \centerline{
   \includegraphics[scale=0.58]{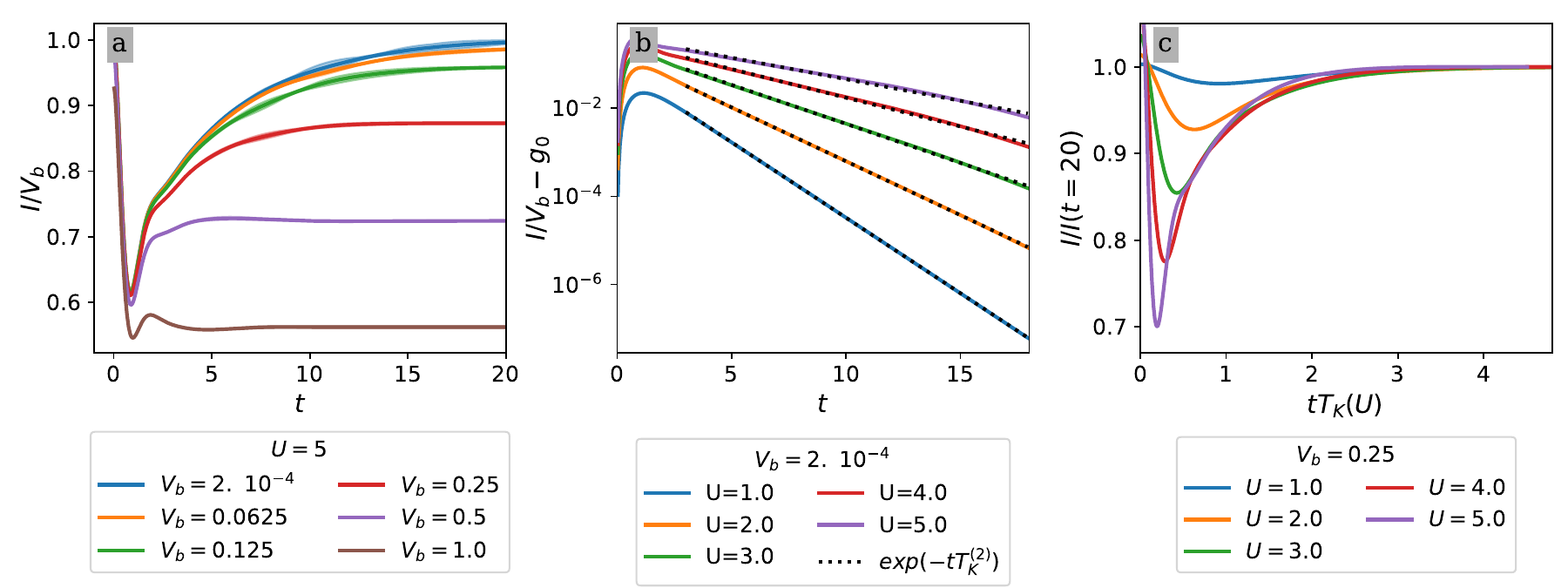}} 
   \caption{
      Kondo scale in current dynamic. 
      \textit{a)} Conductance as a function of time for $U=5$ and by varying $V_b$. 
      \textit{b)} Difference between the conductance and its value at perfect transmission $g_0$ as a function
                  of time $t$ for $V_b=2\ 10^{-4}$ and different values of $U$. Dotted black lines corresponds to linear fits to extract $T_K^{(2)}$. 
      \textit{c)} Current divided by the asymptotic current at $t=20$ as a function of $tT_k^{(2)}$ at $V_b=0.25$ for various $U$.}
\label{fig:Kondo_dynamics} 
\end{figure*}

In order to analyze the formation of the Kondo resonance quantitatively, 
we perform a calculation at very low bias $V_b=2\ 10^{-4}$ 
and study the relaxation of the current to perfect transmission.
The results are shown in Figure
\ref{fig:Kondo_dynamics}. Panel (a) shows $I/V_b$ versus time for different
bias.  Panel (b) shows $I(t)/V_b - g_0$ for the smallest bias voltage. 
After the transient relaxation for $t\sim 1/\Gamma$, 
we observe a clear exponential convergence of this quantity.  
We use this exponential decay to obtain the Kondo temperature $T_K^{(2)}$, 
using a fit $I(t)/V_b - 1\propto \exp(-t T_K^{(2)})$ (dotted lines in panel b). In panel (c), we check that $I/I(t=\infty)$ only depends on $t T_K^{(2)}(U)$ for large $t \gg 1/\Gamma$ for various $U$. 
This scaling property is approximately satisfied at large $U$, in the universal Kondo regime.
Note that in panel (c) the current has been divided by $I(t=20)$ to accounts for a small deviation originating from
the fact that we did not calculate at times large enough to fully reach the perfect transmission.

\begin{figure}
	\centerline{\includegraphics[scale=0.70]{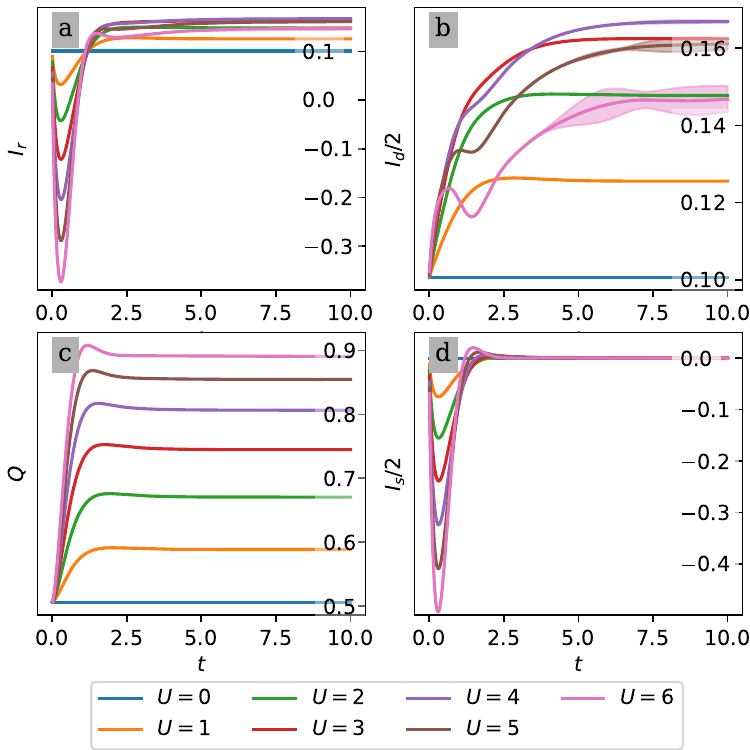}}
	\caption{Dynamic of the current outside particle-hole symmetric regime for $V_b=0.2$ and
	$E_{d}=1$, $\alpha=0.5$. 
   \textit{a)} Current in the right electrode as a function of $t$ for different values of $U$.  
   \textit{b)} Difference of right and left current (divided by two as a function of $t$). 
   \textit{c)} Charge as a function of $t$.
   \textit{d)} Sum of right and left current divided by two.}
	\label{fig:I_with_ed}
\end{figure}

The non-particle-hole symmetric cases illustrate the double time scale relaxation 
even more clearly, as the charge is not fixed ($Q \neq 1$) and has to relax to
its steady state value.  On Figure \ref{fig:I_with_ed}, we plot the relaxation
of the current and charge as a function of time $t$ at $V_b=0.2$ with
$E_{d}=1$ and $\alpha=0.5$.  Here, we distinguish the current from the
left electrode to the dot $I_l$ from the current between the right electrode
and the dot $I_r$ (noted simply $I$ in the rest of this article).  Panel (a)
shows $I_r$ versus time; we observe a strong, fast relaxation over a time scale
$t\sim 1/\Gamma$ followed by a slower and smaller relaxation that is barely
visible.  These large transient currents are due to the fact that the
dot occupation in the long time steady state is very different from its initial
value, hence some current must flow towards the dot to adjust it. This is
illustrated on panel (c) which shows the time evolution of the dot occupation
$Q$. We observe a fast relaxation on the same time scale as the current $I_r$.
Indeed, charge conservation reads $\partial Q/\partial t = - \left( I_{l} + I_{r} \right) \equiv
- I_s$, hence the reorganisation of the charge is linked to the total current
flowing towards the quantum $I_s$. Panel (d) shows $I_s$ versus time: it is
very similar to $I_r$ at small time, then converges to zero after the charge
has relaxed.  In order to isolate the long time relaxation, we
plot the difference $I_d = I_{r} - I_{l} $ in panel (b).
$I_d$ exhibits the long time relaxation associated to the formation
of the Kondo resonance and it converges asymptotically to $I_d = 2 I_{r}$.

%% --------------------------------------

\subsection{Magnetization relaxation}
\label{sec:magRelax}

\begin{figure*}
	
	\centerline{
		\includegraphics[scale=0.58]{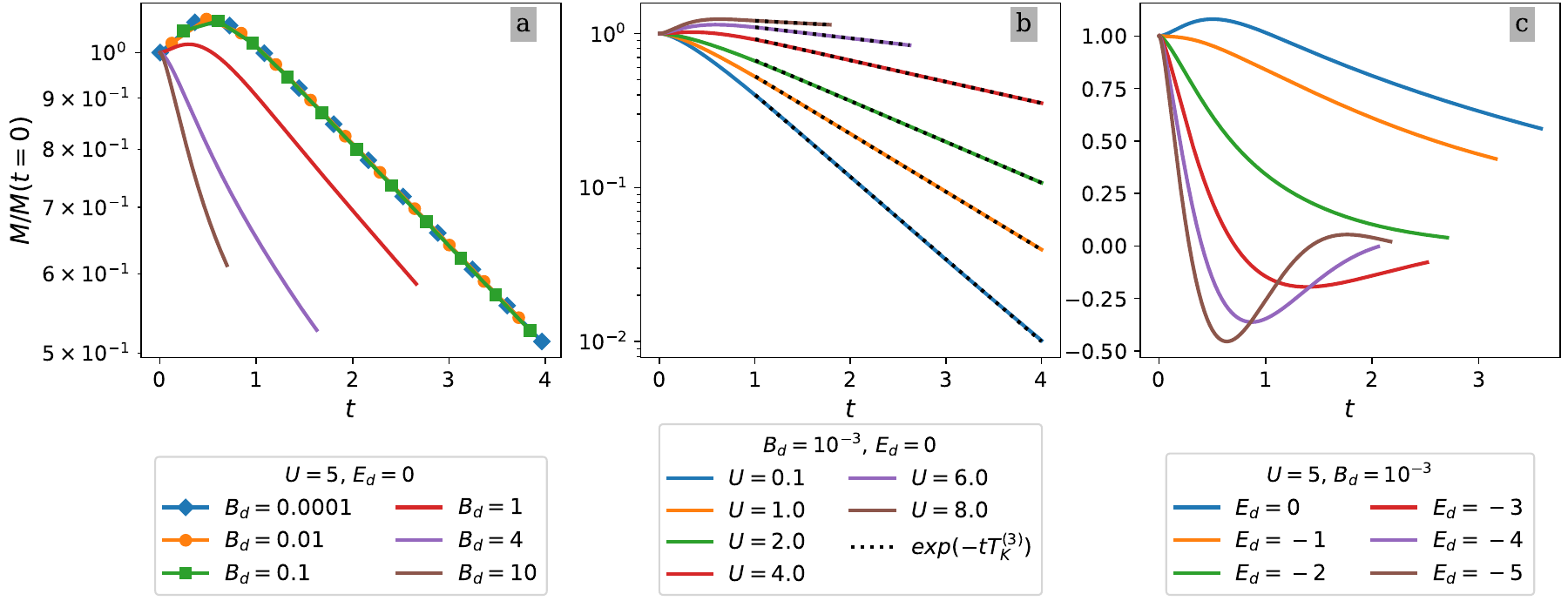}
	}
\caption{
   Magnetization relaxation.
\textit{a)} Magnetization normalized by its initial value as a function of time at $U=5$ for different values of $h$.
\textit{b)}  Magnetization normalized by its initial value as a function of
time at $h=10^{-3}$ for different values of $U$. Black dotted lines corresponds
to linear fits to extract $T_K^{(3)}$. \textit{c)} Normalized magnetization at
$U=5$ and $h=10^{-3}$ with different on-site energy $\epsilon_{d}=-U/2-E_d$.
We use $N=21$ orders.
}
	\label{fig:h_dynamics}
\end{figure*}

Another technique to access the Kondo physics (spin sector) without been
obscured by the charge reorganisation is to
study the relaxation of the magnetization after a quench.
More precisely, we add a small magnetic field 
$B_d(\hat{n}_\uparrow-\hat{n}_\downarrow)$ to induce a small magnetization $M=\langle
\hat{n}_\uparrow-\hat{n}_\downarrow\rangle$. At $t=0$ the interaction is
switched on and the magnetic field removed. 
In practice, we achieve this by using the
following spin-dependent shifting parameters that also set the on-site energy
$\epsilon_d=-\frac{U}{2}$ to the particle-hole symmetry point: 
\begin{equation}
\alpha_{\sigma} = \frac{1}{2}\left( 1 + (-1)^\sigma \frac{2 B_d}{U}\right)
\end{equation}
The results are presented in Fig.~\ref{fig:h_dynamics}. 
First, we check in  Fig.~\ref{fig:h_dynamics}a that we are in linear response regime
for $B_d\ll T_K$. Second, in this low field regime, the relaxation of
the magnetization is well fitted by an exponential $M(t) \propto \exp(-t
T_K^{(3)})$ after a rapid transient regime, Cf  Fig.~\ref{fig:h_dynamics}b.
Note that these data correspond to the
bare calculation without any cross-extrapolation. Indeed, for this case we have
observed that the function $M(t,U)$ does not exhibit the low rank structure
displayed by other quantities and that permits the cross-extrapolation
\cite{Jeannin24}.

Finally, on Fig.~\ref{fig:h_dynamics}c, we change the value of $E_d$ away from the particle-hole
symmetric point.
The magnetization does not exhibit the simple exponential relaxation observed for $E_d=0$. 
It has oscillations as a function of time $t$, due to the fact that
the charge is no longer fixed (as explained in \cite{Haldane78} charge
fluctuation effects become non-negligible).
Nevertheless, for intermediate values of $E_d$ it is still possible to extract a Kondo temperature.

%%%%%%%%%%%%%%%%%%%%%%%%%%%%%%%%%%%%%%%%%%%%%%%%%%%%

\section{Kondo temperature}
\label{sec:Kondo}

The emergent Kondo temperature, $T_K$, is the sole relevant energy scale in the
low-temperature, universal regime of the Anderson model. As discussed in
previous sections, $ T_K $ can be extracted from various quantities: the width
$ T_K^{(1)} $ (at half-maximum) of the Kondo peak in the differential
conductance $ \partial_{V_b} I(V_b) $ (see Fig. \ref{fig:cur_I_Vb}d), the
relaxation time $ 1/T_K^{(2)} $ associated with the current (cf. Section
\ref{sec:currentRelax}), and the relaxation time $ 1/T_K^{(3)} $ of the
magnetization (cf. Section \ref{sec:magRelax}). Note that $T_K$ is \emph{not}
an universal quantity, its precise value depends on how the details of the
model at high energy. Hence we do not expect e.g. an exact match with values
found using Bethe Ansatz. Also the universal regime is only expected when 
$T_K \ll \Gamma$, \cite{Hewson93,Andrei95}.

\begin{figure}
\centerline{ \includegraphics[width=0.48\textwidth]{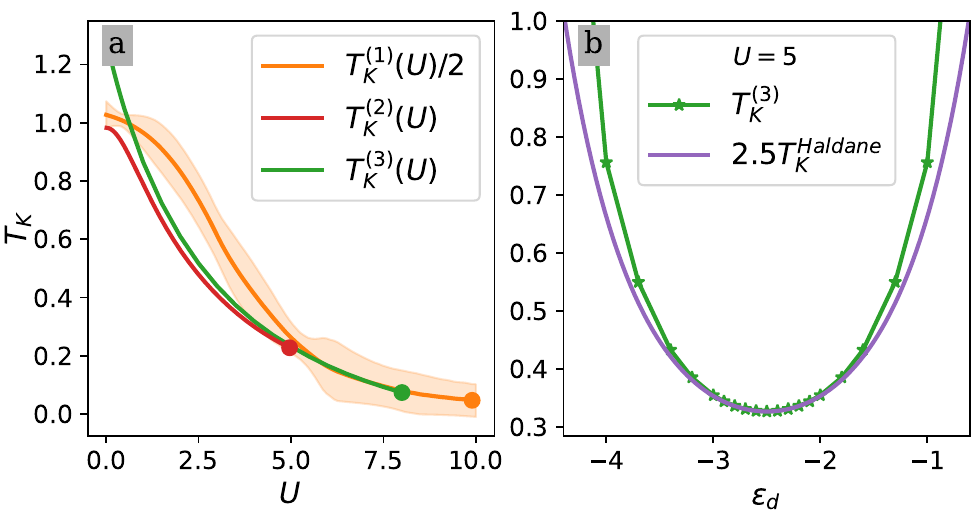} }
\caption{Kondo temperature. \textit{a)}
	Comparison of the Kondo temperature obtained with the width of the Kondo conductance peak $T_K^{(1)}$, the relaxation of the current toward perfect transmission at small bias $T_K^{(2)}$ and the relaxation of the magnetization $T_K^{(3)}$, the points indicate largest value obtained. 	\textit{b)} Kondo temperature $T_k^{(3)}$ and $T_K^{\text{Haldane}}$ as a function of $\epsilon_d$ at $U=5$.}
	\label{fig:Kondo_temperature}
\end{figure}

As shown in Fig. \ref{fig:Kondo_temperature}a, our different definitions of Kondo temperature exhibit good agreement, differing only by a multiplicative factor for values of $ U > 5 $, when the dot is in the universal Kondo regime. 
Outside the particle-hole symmetric point, we observe that $T_K^{(3)}$ varies quadratically with $\epsilon_d$ around $\epsilon_d = -U/2$, in agreement with the predictions of analytical expressions derived by Haldane in \cite{Haldane78,Krishna-murthy75,Krishna-murthy80}. The Haldane expression holds when $U \gg \Gamma$ and is given by:
 \begin{equation}
T_K^{\text{Haldane}}(U) = \frac{1}{2\pi} e^{1/4 + \gamma} \sqrt{\frac{2\Gamma U}{\pi}} e^{\frac{\pi \epsilon_d (\epsilon_d + U)}{2 \Gamma U}}
\end{equation}
where $\gamma$ is the Euler's constant. We find good agreement with this expression for 
$U>4$ (up to a multiplicative factor) but it should not be used for smaller values of $U$ where it is no longer accurate.

%=========================================================

\section{The Coulomb-Blockade regime}
\label{sec:coulomb-blockade}

In this section, we focus on the dot occupation for relatively large values of $U$
and study the deviations from the standard theory of Coulomb-blockade

\subsection{Comparison with sequential tunneling theory}

\begin{figure}
\centerline{
\includegraphics[scale=0.70]{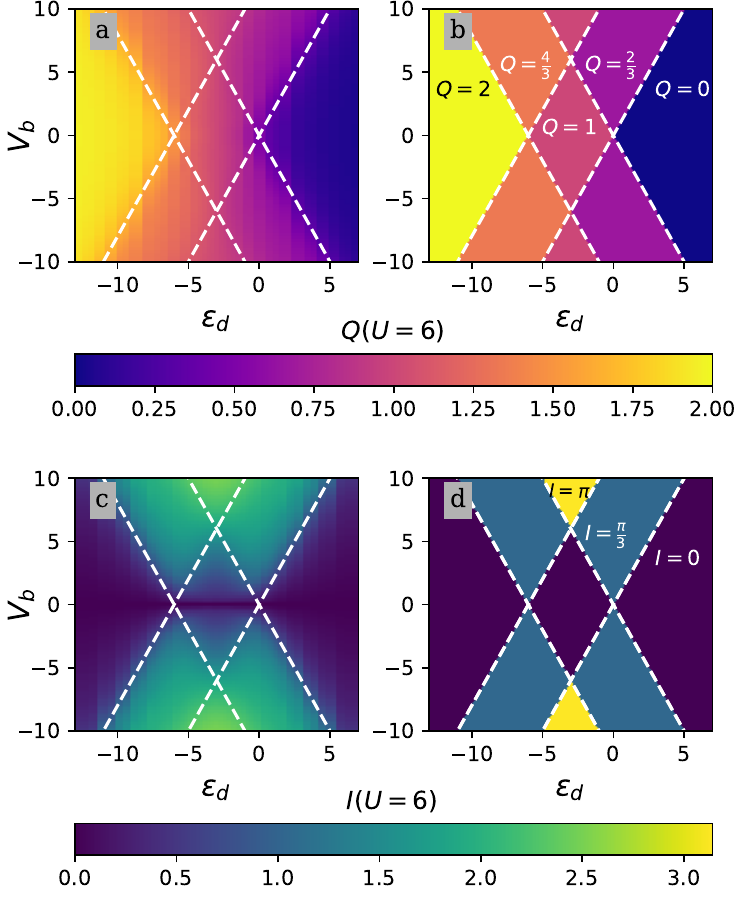}
}
\caption{Comparison with semi-classical model of appendix \ref{sec:semi-classical}. \textit{a)} Charge at $U=6$ and long time as a function of $\epsilon_d$ and $V_b$. White dashed lines corresponds to the semi-classical model. \textit{b)} Semi-classical approximation of plot a), see section \ref{sec:semi-classical}. \textit{c)} Current at $U=6$ and long time as a function of $\epsilon_d$ and $V_b$. \textit{d)} Semi-classical approximation of c). $N=16$.
}
\label{fig:classic}
\end{figure}

Ignoring the charge and spin quantum fluctuations, the charge and current of
the Anderson model can be calculated from the sequential tunneling theory that
describes the tunneling process through the dot as independent stochastic
incoherent events \cite{Beenakker91}. 
The associated simple analytical expressions are derived in Appendix \ref{sec:semi-classical} for completeness. 
Figure \ref{fig:classic}, shows a comparison between the sequential tunneling
calculation and the results obtained with the complete many-body simulation at
$U=6$, for the charge and the current. In the $(\epsilon_d, V_b)$ plane, both
calculations predict variations in the form of staircases: this phenomenon is
known as Coulomb diamonds and has been repeatedly observed experimentally
\cite{Mak13,Dekker99,Hofheinz06,Stampfer08,Leturcq09,Kotekar-Patil19}. 
The exact calculation does indeed match the sequential tunneling result
qualitatively, despite being somewhat blurred compared to the sharp feature of
the classical approach that ignores quantum fluctuations. This agreement is
only qualitative and somewhat deceptive however: it is partly due to the variations
of the charge being very large on the scale of the colormap.
Indeed, for instance, the V-shape feature discussed earlier in the context of
Fig.\ref{fig:charge_Vb_ed0} is barely visible on the colorplot.  

% ----------------------------------------------------

\subsection{Coulomb Blockade at large $U$}
\label{sec:coulomb_blockade}

\begin{figure*}
\centerline{	
 \includegraphics[scale=0.6]{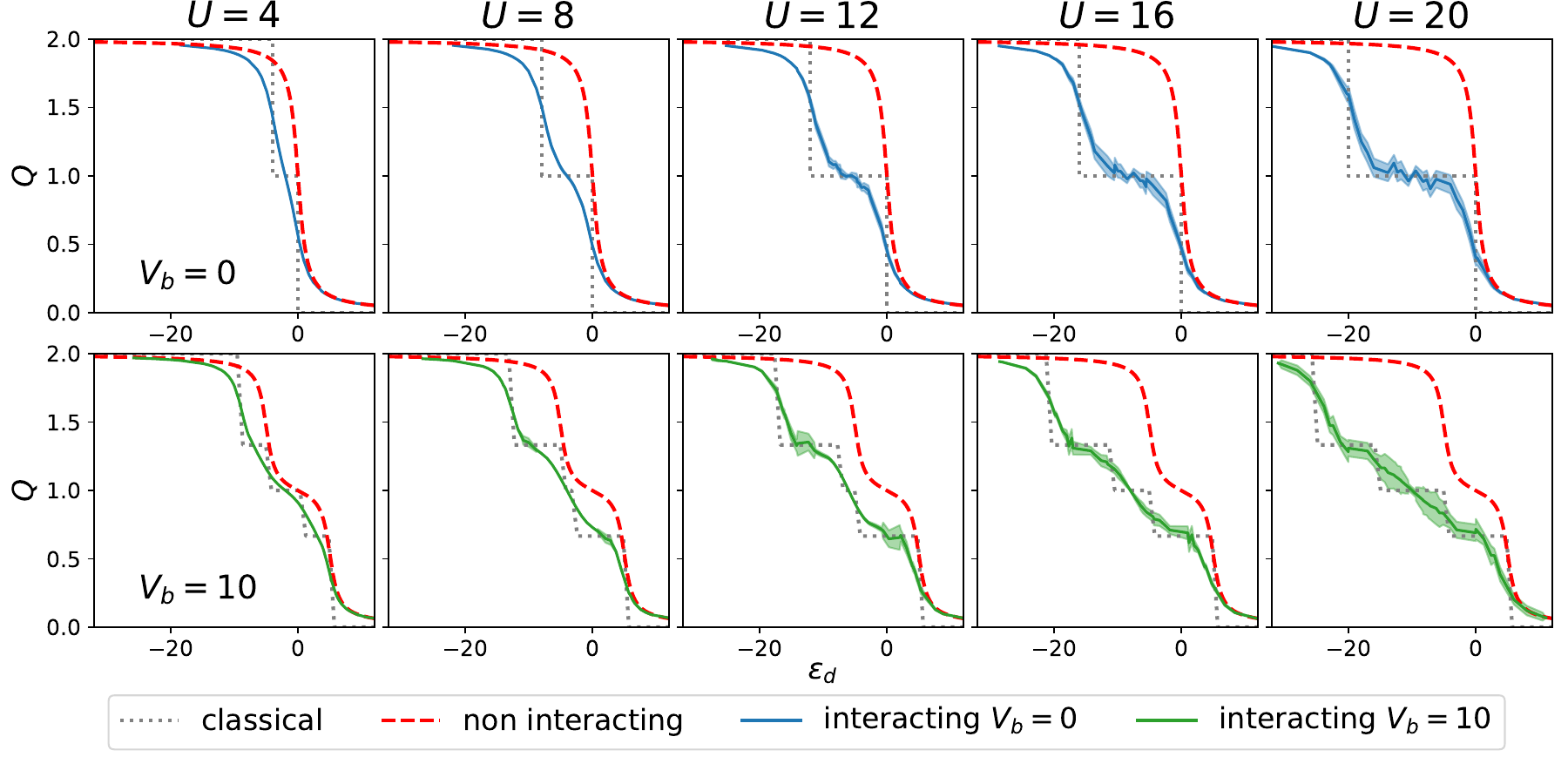}
}
\caption{Effect of the quantum fluctuations in the "Coulomb staircase". Charge at $t=5$ as a function of $\epsilon_{d}$ for different values of $U$. The red line indicates non interacting model, the grey dotted line the classical model, the blue curve the interacting model with the method presented. Dashed part of the curve corresponds error bars. $N=23$}
\label{fig:coulomb_blockade}
\end{figure*}

Figure \ref{fig:coulomb_blockade} shows the charge as a function of
$\epsilon_d$ for different values of $U$ using $V_b=0$ (upper panels) and
$V_b=10$ (lower panels). For comparison, the sequential tunneling limit is
shown as a grey dotted line, and the exact non-interacting limit $U=0$ model as
a red dashed line. 

Remarkably, very high values of $U/\Gamma \ge 20$ are necessary to observe a
relatively sharp Coulomb staircase, i.e. to essentially suppress charge
fluctuations and enter the regime where the Anderson model is well described by
its Kondo model limit. On the other hand for such a large value of $U$, $T_K$
is very small: $T_K/\Gamma \approx 5.\ 10^{-4}$. Suppose that in order to
observe a clear Kondo ridge, one works at a temperature $T \approx T_K/10$.
For a base temperature of $T= 10 mK$ (a typical dilution fridge temperature),
one obtains $T_K = 100 mK$, hence $\Gamma \approx 200 K$ and $U\approx 4 000K$.
On the other hand, typical experimental values of the charging in semiconductor
structure are of a few Kelvins or a few tens of Kelvin at most. Hence, we conclude
that in the actual observation of Kondo physics, one is almost \emph{never}
in the limit of well defined dot occupation and one must use the full Anderson
model rather than its (slightly more tractable) Kondo limit. A notable exception to the above are molecular quantum dots - of atomic sizes - where a much higher charging energy, up to $10^4$K, could be achieved \cite{Park02,Roch08,Roch09}.

These considerations might be of interest for quantum bit physics. Indeed,
coupling two spin qubits with a two qubit gate requires to be able to tune the
system to have sufficient tunneling between the two dots. For systems where the
Coulomb staircases are extremely sharps and (to the best of our knowledge)
Kondo physics is never observed such as Silicon MOSFETS, it is reasonable to
expect that the absence of a large dot-lead tunneling regime will also
correspond to an absence of a large dot-dot tunneling regime, hence to face
difficulties to achieve reliable and controllable two qubits gates.

Out-of-equilibrium,  for large bias $V_b=10$, we observe that  two plateaus
appear for charge values equal to $Q=2/3$ and $Q=4/3$ in agreement with the
sequential tunneling limit. Note however that the plateau at $Q=1$ predicted by
this theory is not clearly visible in the exact solution suggesting that quantum
fluctuations of the charge are relevant in this regime. Besides, both in 
equilibrium and out-of-equilibrium, the formation of
the plateau with $\epsilon_d$ is smoother than in the
semi-classical model, due to higher order tunneling effects.

%=================================

\section{Conclusion}

In this article, we have computed the ``Coulomb diamonds", the ``Coulomb
staircase" and the ``Kondo ridge" with a numerically exact technique. The
technique is exact in the sense that it gives controlled error bars that can be
systematically improved by adding computing time. These results open the
possibility of a systematic comparison with experiments, out-of-equilibrium,
that is not impaired by the ubiquitous set of uncontrolled approximations that
one must usually resort to in order to solve out-of-equilibrium many-body
problems. The approach is directly extensible to more complex situations such
as double or triple dots, non-trivial environments (e.g. a dot inside one arm
of an interferometer), other quantities (e.g. Seebeck effect) and
time-dependent perturbations (work in progress).

We have used the TTD  and the cross-extrapolation methods to obtain 
a comprehensive benchmark of the SIAM model, scanning a large range
of gate potential $\epsilon_d$ and bias voltage $V_b$. We made quantitative contacts
with a number of previous works.

On the technical level, we have demonstrated that the cross-extrapolation method is robust, generating reliable error bars in all cases, and enabling the computation of a comprehensive phase diagram.
However, we also discovered a physical quantity (the magnetization) for which
the low-rank assumption is not satisfied, contrary to the charge or the current.
It is an interesting question to try to understand which quantities will be 
amenable to a low-rank extrapolation.

%=================================

\acknowledgements 

The authors thank Serge Florens for useful discussions and insightful remarks.
 
%\newpage
%\clearpage   
\bibliography{biblio_SIAM}

\appendix

\section{Master equation for Anderson model}
\label{sec:semi-classical}
In this section one describes a simple semi-classical form of the current flowing through a quantum dot following the standard theory of sequential tunneling \cite{Beenakker91}. We describe the state of the dot by its probability to be occupied by zero ($p_0$), one 
($p_{\uparrow}$ and $p_{\downarrow}$ ) and two electrons ($p_{\uparrow \downarrow}$) with
$p_0+p_{\uparrow}+p_{\downarrow}+p_{\uparrow \downarrow}=1$. The tunneling events are treated as independent stochastic process with a rate for electron entering 
($\bar{\Gamma}(E)$) and leaving ($\tilde{\Gamma}(E)$) the dot being respectively,
\begin{align}
\bar{\Gamma}(E) &= \Gamma_L f_L(E)+\Gamma_R f_R(E)\\
\tilde{\Gamma}(E) &= \Gamma_L \left[1-f_L(E)\right] + \Gamma_R \left[1-f_R(E)\right]
\end{align}
where $E$ is the difference of energy of the dot between after and before the tunneling event. For example the probability for an electron to leave the dot towards the left electrode when it is in state $\uparrow$ is given by $p_\uparrow \Gamma_L (1-f_L(\epsilon_{d}))$ with $\Gamma_L$ the tunnel probability to the left electrode and $1-f_L(\epsilon_{d})$ the probability to have an available state in the left electode. 
This particular process contribute to $\partial_t p_0$. 

In the steady state the corresponding master equation reads,
{\small\par}
\onecolumngrid
\noindent\rule{9cm}{0.4pt}

\begin{equation}
\left(\begin{array}{cccc}
-2\bar \Gamma (\epsilon_{d}) & \tilde{\Gamma}(\epsilon_d) & \tilde{\Gamma}(\epsilon_{d}) & 0 \\
\bar{\Gamma}(\epsilon_{d}) & -\bar{\Gamma}(\epsilon_{d} + U) -\tilde{\Gamma}(\epsilon_{d}) 
 & 0 & \tilde{\Gamma}(\epsilon_d + U)\\
\bar{\Gamma}(\epsilon_{d}) & 0 
& -\bar{\Gamma}(\epsilon_{d} + U) -\tilde{\Gamma}(\epsilon_{d}) & \tilde{\Gamma}(\epsilon_d + U)\\ 
0 & \bar{\Gamma}(\epsilon_{d}+U) & \bar{\Gamma}(\epsilon_{d}+U) & -2\tilde{\Gamma}(\epsilon_{d}+U)
 
 \end{array}\right)
 \left(\begin{array}{c}
 p_0\\p_\uparrow\\p_\downarrow\\p_{\uparrow \downarrow}
 \end{array}\right) = 0
 \label{eq:simple_model}
\end{equation}
\twocolumngrid

\begin{figure}
	
	\centerline{
		\includegraphics[scale=0.45]{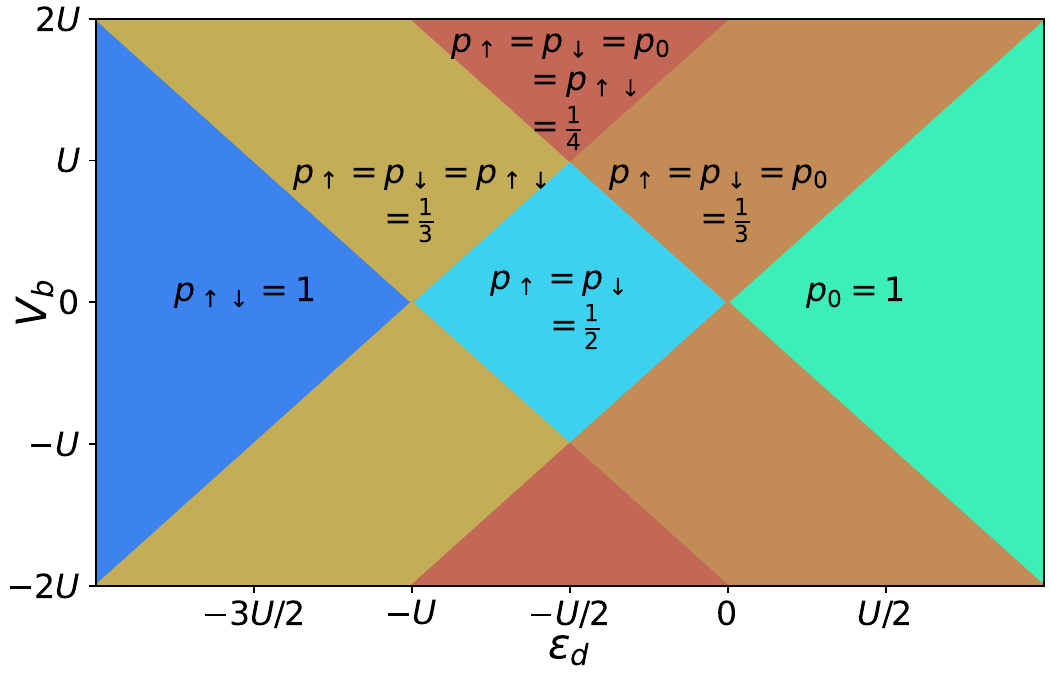}
	}
	\caption{Probability of occupation in the  $(\epsilon_d, V_b)$ plane  for the sequential tunneling semi-classical model. The different separation lines are given by $|V_b|=2|\epsilon_d|$ and $|V_b|=2|\epsilon_d+U|$.}
	\label{fig:classic_model_occupation}
\end{figure}

In this article, we focus on the symetric case $\Gamma_L = \Gamma_R = \Gamma/2$ 
as well as $T=0$ such that $f_{L/R}(E)=1-\theta(E\mp V_B/2)$.  Solving this simple model \eqref{eq:simple_model}, one obtains the different probability values delimited by the different domains in the plane $(\epsilon_d, V_b)$ shown in Fig. \ref{fig:classic_model_occupation}). From the probabilities one can deduce an expectation value for the observables. The charge is given by:
\begin{equation}
	Q= 0\ p_0+1\ (p_\downarrow+p_\uparrow)+2\ p_{\downarrow\uparrow}
\end{equation}
The current flowing from the dot to the left electrode is given by the probability to escape the dot and go in the left electrode minus the probability of going to the dot from the left/right electrode:
\begin{multline}
	I_{l/r}= \Gamma/2[(p_ \uparrow+p_ \downarrow)(1-f_{L/R}(\epsilon_d)) + 2p_ {\uparrow \downarrow} (1-f_{L/R}(\epsilon_d+U))] \\- \Gamma/2 [2p_0f_{L/R}(\epsilon_{d})+(p_\uparrow+p_\downarrow)f_{L/R}(\epsilon_d+U)]
\end{multline}

%%%%%%%%%%%%%%%%%%%%%%%%%%%%%%%%%%%%%%%%%%%%%%%%%%%%%%%%%%%%%%%%

\section{Example of bare outputs of the TTD technique}
\label{sec:app:coef}

\begin{figure*} 
   \centerline{ \includegraphics[scale=0.49]{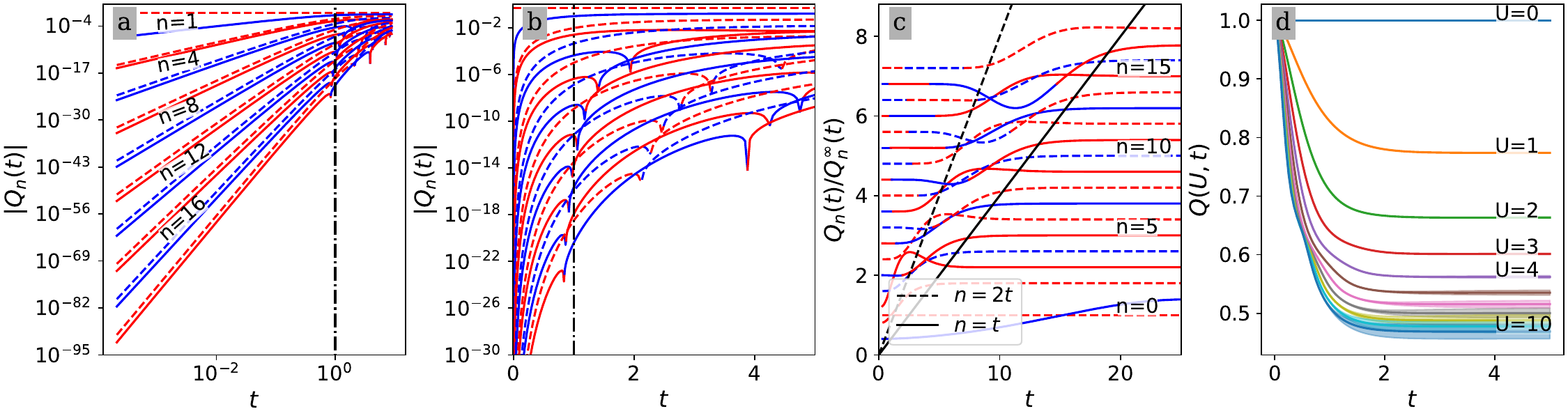} }
   \caption{
      Coefficients of the expansion. \textit{a)} \textit{b)} and
   \textit{c)} absolute value of the coefficients $Q_n$ as a function of time
for the charge using ($\alpha=E_{d}=0$) at three different time intervals. Even
$n$ coefficient are dashed and odd one are plain.  Red (blue) portion of curve
corresponds to positive (negative) values of $Q_n$. The dashed vertical line in
\textit{a)} \textit{b)} indicates $t=1$. In \textit{c)}, the coefficients are
divided by the asymptotical value given by Bethe Ansatz and regularly offset for
readability. The Bethe ansatz value is reached assymptotically with high precision. The lines $n=t$ and $n=2t$ are guide to the eye. \textit{d)} Charge as a
function of time for the same parameters and different values of $U$, the shaded regions
corresponds to error bars.
}\begin{flushleft}

\end{flushleft}
\label{fig:coef}
\end{figure*}

Figure \ref{fig:coef} shows an example of the bare data coming out of the
tensor train diagrammatic technique: the coefficients $Q_n(t)$ as functions of
time in double log scale (panel a) simple log scale (panel b) and normal scale
(panel c). The even coefficients are shown with dashed lines and the odd
coefficients with solid lines. When the coefficients are positive (negative)
the curve is red (blue). A striking feature of this figure is the ability of
the technique to calculate values that span almost \emph{one hundred} orders of
magnitude (see the y-axis of panel a). As advertised, $Q_n(t)$ can be in three
different regimes (panel c): perturbative for $t\ll n/\Gamma$, stationary for
$t\gg n/\Gamma$ and transient for $t\sim n/\Gamma$. Panel (d) shows the dot
occupation as a function of time calculated from these coefficients: it is
remarkable to observe that despite the numerous oscillations of the
coefficients, due to internal compensations, the charge evolves smoothly with
time.

%%%%%%%%%%%%%%%%%%%%%%%%%%%%%%%%%%%%%%%%%%%%%%%%%%%%%%%%%%%%%%%%

\section{Additional benchmarks with previous literature}
\label{sec:app:bench}

\begin{figure}
	
	\centerline{
		\includegraphics[scale=0.45]{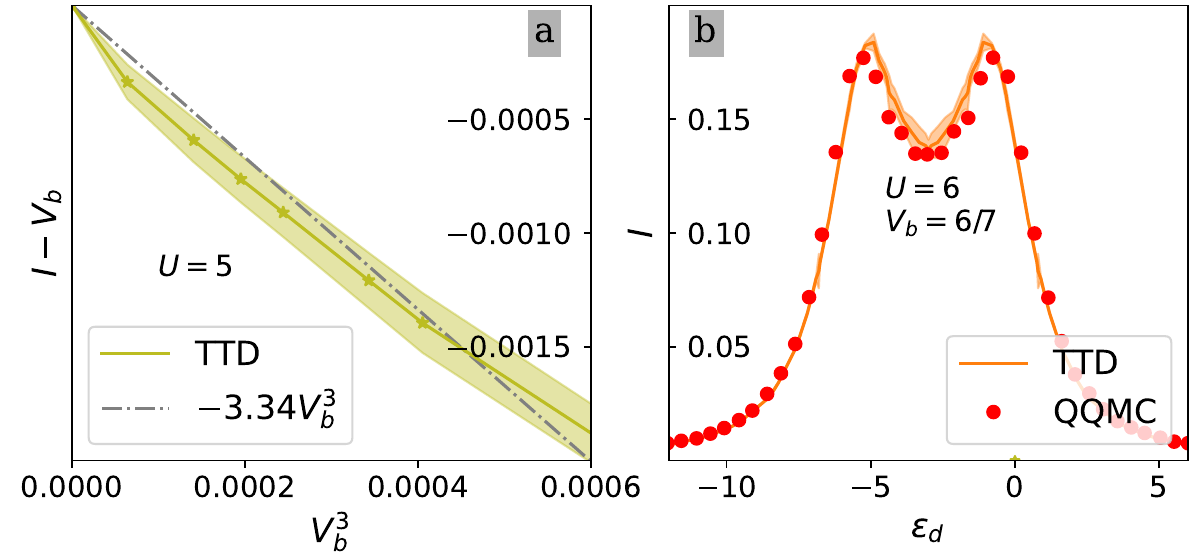}
	}
	\caption{Comparison of TTD with Fermi-liquid theory and QQMC. Panel a $I-V_b$ as a function of $V_b^3$ ($I$ is $\frac{2e}{h\Gamma}$ units). The green line is the calculation obtain with our method, the dashed grey line is obtained from Fermi-liquid theory. Panel b, current at $U=6$ and $V_b=6/7$ as a function of $\epsilon_d$ comparison of our method (solid line) with QQMC (points)  }
	\label{fig:benchmark}
\end{figure}

Figure \ref{fig:benchmark} shows additional comparison of our results with two existing benchmarks available in the litterature. 
\cite{Mora15} calculated the $O(V_b^3)$ deviation to the linear conductance using
a Fermi liquid theory to relate this deviation to the equilibrium properties of the dot obtained from Bethe ansatz exact calculations. This results is interesting because there are very few exact results out-of-equilibrium. \cite{Mora15} predict that at low bias one should observe,
\begin{equation}
\frac{I(U,V_b)}{g_0} = V_b-V_b^3\left(\alpha_1(U)\frac{2}{12}+\phi_1(U)\frac{5}{12}\right)+o(V_b^3)
\end{equation}
where the two coefficients $\alpha_1$ and $\phi_1$ can be obtained from the Bethe ansat equations. We consider the case $U=5$ for which $\alpha_1(U=5) \approx 2.74$ and $\phi_1(U=5) \approx 2.55$, so that in the chosen units
\begin{equation}
I(U=5,V_b) \approx V_b-3.34 V_b^3 + o(V_b^3)
\end{equation}
The comparison between this expression (dotted-dashed) and our results (full line) is shown in panel (a) and is indeed quantitative for small biases. For values of $V_b>0.05$, the third order theory is not sufficient and higher orders need to be added. Note that obtaining these results require a very high accuracy as this is a tiny deviation to the main linear contribution.

Panel (b) corresponds to the current at $U=6$ as a function of $\epsilon_d$ for a bias of $V_b=U/7$. We compare the results obtained (solid line) with those of quasi-quantum Monte Carlo \cite{Macek20}. Both results are in quantitative agreements within their respective  error bars.

%%%%%%%%%%%%%%%%%%%%%%%%%%%%%%%%%%%%%%%%%%%%%%%%%%%%%%%%%%%%%%%%
\section{Cross Extrapolation results}
\label{sec:app:CrossExtrapol}

\begin{figure}
	
	\centerline{
		\includegraphics[scale=0.70]{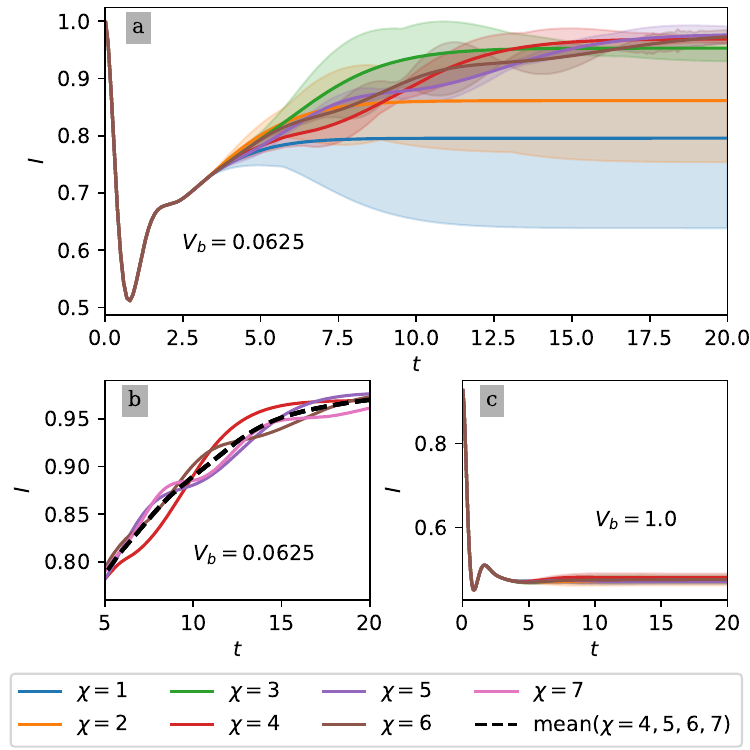}
	}
	\caption{Errors on the cross-extrapolation at small bias. \textit{a)} current at $U=6$ and $V_b=0.0625$ for different values of $\chi$ the rank of the cross-extrapolation as a function of time, estimated error bars are given by shaded ereas sections. \textit{b)} current at $U=6$ and $V_b=0.0625$ for $\chi=4,5,6,7$ as a function of time. To remove oscillations, it is possible to average over several $\chi$ values (black dashed line). \textit{c)} same as \textit{a} for $V_b=1$, error bars are optimal for smaller ranks on the order of $\chi=1,2$
	}
	\label{fig:rank_Kondo}
\end{figure}

Most of the results of this article have been obtained by extending the bare data to larger values of $t$ and $U$ using the cross-extrapolation technique \cite{Jeannin24}. 
The technique exploits the fact that, in almost all the cases that we have studied, an
observable $Q(t,U)$ has a small rank $\chi$, i.e. can be written as the sum over few products of function 
\begin{equation}
Q(t,U) \approx \sum_{i=1}^\chi f_i(t) g_i(U)
\end{equation}
Usually, we have found that a low value $\chi = 2-3$ was sufficient for the extrapolation.
In one circonstance, however, a higher value was required.
This case, the current at very small bias voltage inside the Kondo ridge, is shown in Figure \ref{fig:rank_Kondo}. We attribute this increase of rank to the onset of the Kondo effect that makes the time dynamics non-trivial. 
Figure \ref{fig:rank_Kondo} shows that, at small $V_b$ (panel a), a rank $\chi\ge 4$
is necessary for convergence. Note that the results for lower values of $\chi$ are still correct, the error bars are simply much larger.
For $\chi\ge 4$, we observe that the extrapolations contain some spurious oscillations,
compatible with our error bars. For this particular regime (and only in this case) we have used the average over $4\le\chi\le 7$ as our reference extrapolation (panel b). 
The (now removed) spurious oscillations are of the same order as the calulated error bars. 
In contrast, for the case of larger $V_b$ (see panel c), the error is not significantly smaller for large values of $\chi$ and the extrapolation is more straightforward. 

%%%%%%%%%%%%%%%%%%%%%%%%%%%%%%%%%%%%%%%%%%%%%%%%%%%%%%%%%%%%%%%%
\section{Role of the $\alpha$ shift in series convergence}
\label{sec:app:alphaRole}

To reach a given target in $(\epsilon_d,U)$, the TTD technique proposes different possible expansions by playing with different combinations of $E_d$ and $\alpha$, i.e. with different ways to split the Hamiltonian between initial state and perturbation.
For different values of $(E_d,\alpha)$ such that $\epsilon_{d} = E_d-\alpha U $ the steady- state is identical but the dynamic is different. The convergence radius of the stationary series as well as the easiness with which one can extrapolate the steady-state depends on a a judicious choice of parameters $(\alpha,E_d)$ \cite{Profumo15}.  We find empirically that aiming for the largest radius of convergence is a good strategy to facilitate the cross-extrapolation.

\begin{figure}
	
	\centerline{
		\includegraphics[scale=0.51]{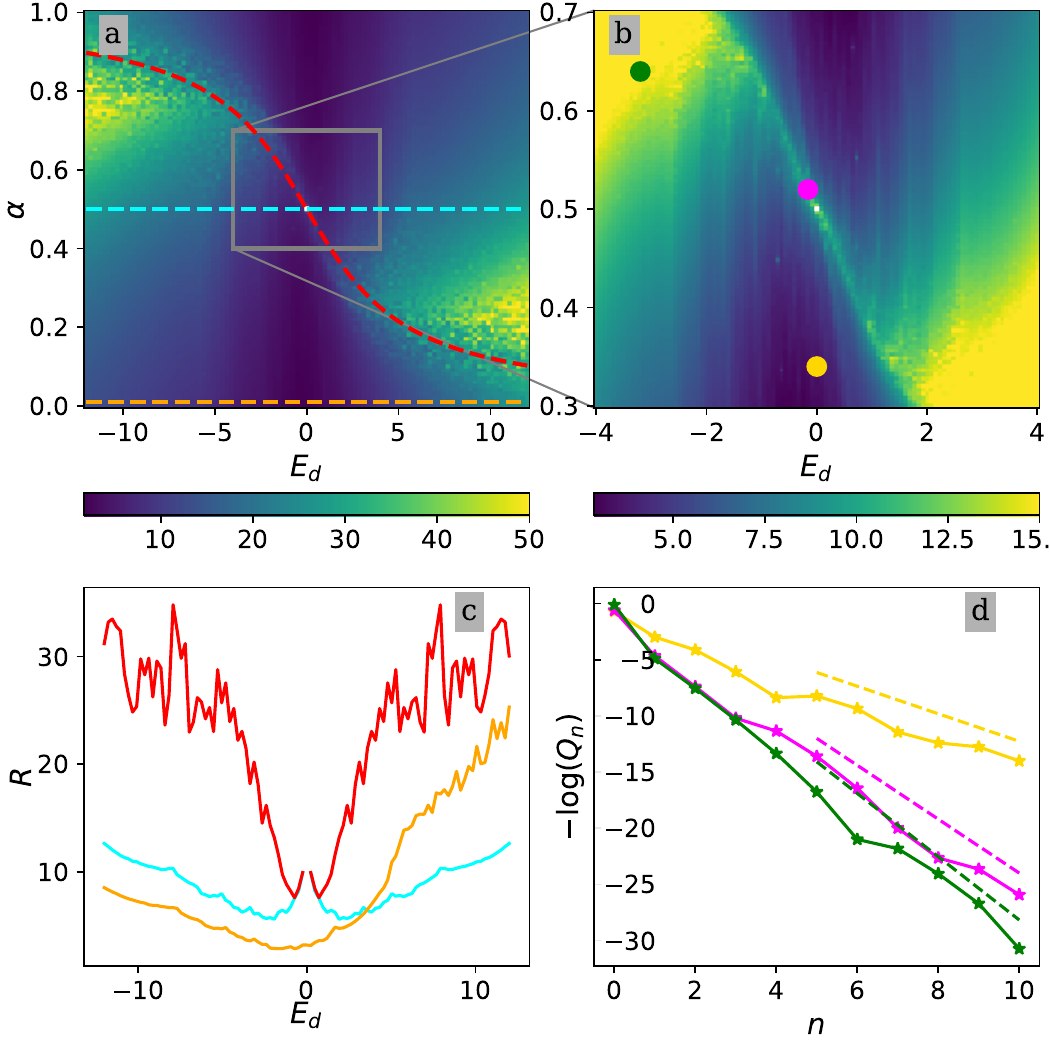}
	}
	\caption{Asymptotic radius of convergence $R$ for the charge $Q$ in the dot.  \textit{a)} color plot of the asymptotic radius of convergence of the series $Q_n$ at infinite time as a function of $\alpha$ and $E_d$. \textit{c)} asymptotic radius of convergence $R$ along the differents cuts in plot a. \textit{b)} zoom of plot \textit{a}. \textit{d)}  $-\log(|Q_n|)$ as a function of $n$ for values of $\alpha$ and $E_d$ indicated in c. Dotted lines correspond to the fit $\sim 1/R^n$ from which we estimate $R$.
	}
	\label{fig:radius_conv}
\end{figure}

This radius of convergence $R$ at large time (formally $t\rightarrow \infty$, in practice
$t\gg n$) is extracted by noting that the series behaves as $Q_n \sim 1/R^n$ and considering a robust linear regression on the logarithm of the coefficients (mediane of the linear regression for different subset of points). 
Figure \ref{fig:radius_conv} Panel (d) indeed shows the logarithm of the absolute values of the coefficients $Q_n$ as a function of $n$ for the different tuples of $(\alpha,E_d)$ indicated by circles in panel (b). The dotted lines correspond to the fits $Q_n\sim 1/R^n$.

Figure \ref{fig:radius_conv}, panel (a)  (zoomed in b) shows a colormap of the radius of convergence $R$ of the long-time equilibrium charge series as a function of $\alpha$ and $E_d$. The yellowish region in panels (a,b) corresponds to the largest radius of convergence for which we could reach the highest values of $U$. 
Panel (c) shows cuts along the plot $(E_d,\alpha)$ (dashed lines in panel a). It is clear that there is an optimal choice of parameters to optimize $R$, it is given by the dashed red curve of panel (a) which we have used in most of our calculations.

\begin{figure}
	
	\centerline{
		\includegraphics[scale=0.5]{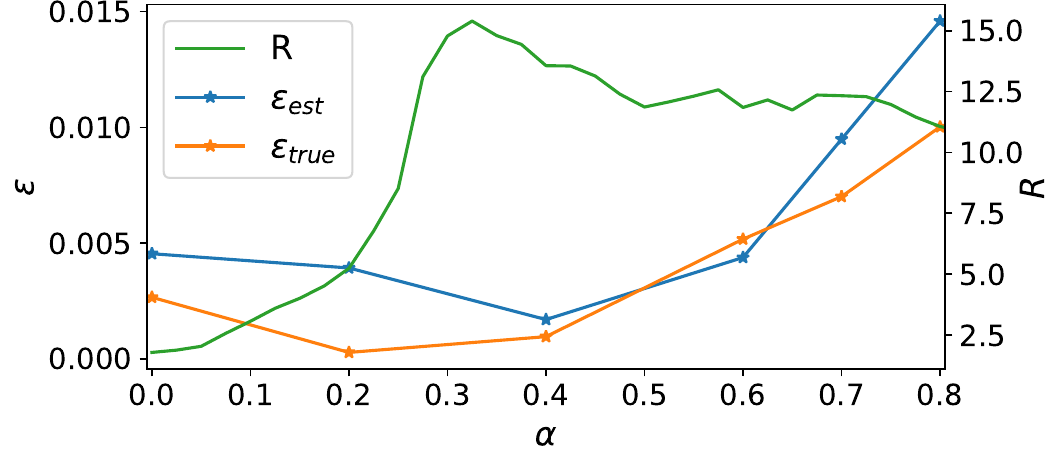}
	}
	\caption{Error and asymptotic radius of convergence. Left axis: estimated error (blue) and true error (orange) on the $Q(U=8)$ for long times as a function of the shift $\alpha$ and fixing $E_d$ such that $\epsilon_d = E_d-\alpha U = 0$. The reference value is given by Bethe ansatz. Right axis, radius of convergence of the asymptotic series as a function of $\alpha$. }
	\label{fig:opt_alpha}
\end{figure}

Figure \ref{fig:opt_alpha} shows the extrapolated error versus $\alpha$ for $\epsilon_d = E_d-\alpha U \equiv 0$ and $U=8$. After extrapolation, at long time, the results are all in agreement but the accuracy of the results depends on the quench considered. For a fixed value of $(U,\epsilon_d)$, we find that the optimal choice of $(\alpha,E_d)$ corresponds to selecting the largest value of $R$.

%%%%%%%%%%%%%%%%%%%%%%%%%%%%%%%%%%%%%%%%%%%%%%%%%%%%

\section{Additional data on charge relaxation}
\label{sec:chargeRelax}

\begin{figure*}
\centerline{
\includegraphics[scale=0.65]{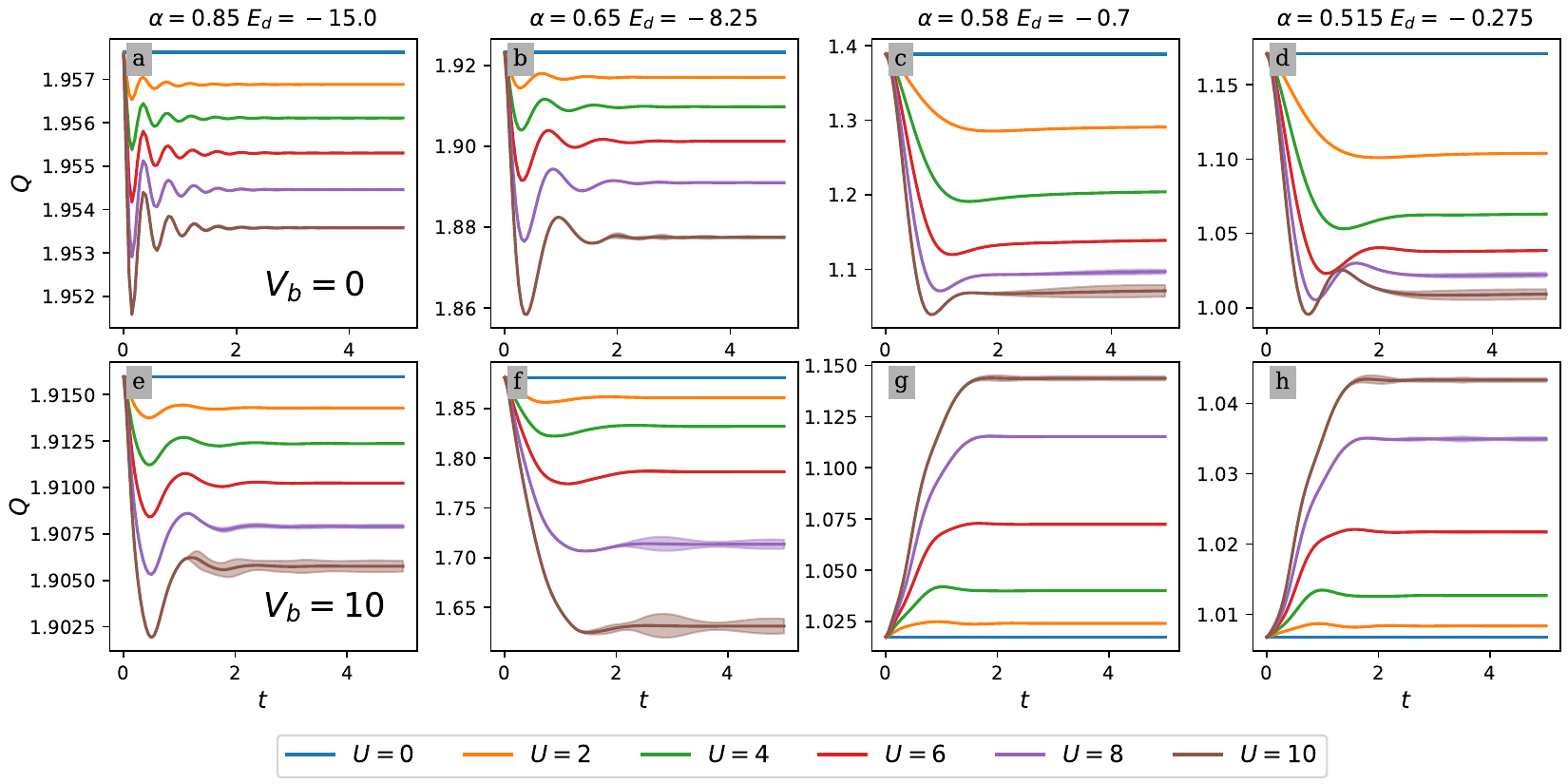}
}
\caption{Charge relaxation. Charge is plotted as function of time for different
values of $U$ using $N=23$ coefficients, different $(\alpha,E_d)$ are
considered. The first line corresponds to $V_b=0$, the second one to $V_b = 10$}
\label{fig:some_squench_Q_u_t}
\end{figure*}

Figure \ref{fig:some_squench_Q_u_t} shows the evolution of the charge as a
function of time and $U$ for different values of the parameters $(E_d,\alpha)$.
The upper panels corresponds to $V_b=0$, the low panels to $V_b=10$. These
quenches are those used to determine the asymptotic value of the charge shown
in Figure \ref{fig:coulomb_blockade}.  One observes a fast oscillatory
relaxation of the charge on a time scale $t \sim \Gamma^{-1}$. The oscillations
are damped in presence of a large bias voltage.

\section{Relation between Lesser Green function and physical observables} 
\label{sec:Lesser Green function and observables}

The TTD method produces a perturbative expansion of the Keldysh lesser Green's function, which serves as the foundation for calculating physical observables. The lesser Green's function is defined for temporal arguments $t,t'$, spatial indices $i,j$, and spin indices $\sigma,\sigma'$ as:
\begin{equation}
G_{i\sigma,j\sigma'}^<(t,t') = i \langle c_{j\sigma'}^\dagger(t') c_{i\sigma}(t) \rangle.
\label{eq:lesser_def}
\end{equation}

The central quantity of interest is the equal-time Green's function, which admits a series expansion in powers of the interaction strength $U$:
\begin{equation}
G_{i,j}^<(t,t) = \sum_{n=0}^\infty [G_{i,j}(t)]_n U^n,
\label{eq:lesser_expansion}
\end{equation}
where $[G_{i,j}(t)]_n$ denotes the $n$-th order coefficient. 

For the dot (position index 0), spin degeneracy reduces the total charge to:
\begin{equation}
Q(t) = -2i G_{0,0}^<(t,t),
\label{eq:charge}
\end{equation}
where we suppress spin indices due to identical behavior for both spin species. Substituting the expansion \eqref{eq:lesser_expansion} yields the perturbative charge coefficients:
\begin{equation}
Q_n(t) = -2i [G_{0,0}(t)]_n.
\end{equation}

Current operators are derived from the time evolution of charge, governed by the Heisenberg equation of motion:
\begin{equation}
\frac{d}{dt}\langle c_0^\dagger(t)c_0(t)\rangle = i\langle [H_0(t), c_0^\dagger(t)c_0(t)] \rangle.
\end{equation}
Considering the first site of the left electrode ($i=-1$) and of the right electrode ($i=1$), this yields to:
\begin{equation}
\frac{dQ(t)}{dt} = -2\gamma\sum_{s=\pm1}\left[G_{s,0}^<(t,t) - G_{0,s}^<(t,t)\right].
\label{eq:charge_current_rel}
\end{equation}

The right-lead current operator (positive outflow convention) is identified from \eqref{eq:charge_current_rel} as:
\begin{equation}
I_r(t) = 2\gamma\left[G_{1,0}^<(t,t) - G_{0,1}^<(t,t)\right].
\label{eq:green_current_def}
\end{equation}
Substituting the perturbative expansion \eqref{eq:lesser_expansion} into \eqref{eq:green_current_def} produces the current coefficients:
\begin{equation}
I_n(t) = -4\gamma\,\mathrm{Re}\left([G_{0,1}(t)]_n\right).
\label{eq:green_current}
\end{equation}

\section{Non-interacting Green's function}
\label{sec:Non-interacting Green}

In this section, we derive
the non-interacting lesser and greater  Green function of the SIAM model. These functions form the input of the TTD algorithm, it is important to evaluate them both precisely and quickly. For this article, the problem is sufficiently simple so that we can calculate these
propagators analytically.

\subsection{Retarded Green's function of semi-infinite lead}
We first calculate the retarded Green's function $g_{\text{lead}}^r(\omega)$ for a semi-infinite tight-binding lead with hopping amplitude $\bar{\gamma}$. Using the infinite system's translational invariance, we consider adding one site to the lead's end. The Dyson equation yields the recursion relation (\cite{Waintal24} Eq.[125]):

\begin{equation}
g_{\text{lead}}^r(\omega) = \frac{1}{\omega- \bar{\gamma} g_{\text{lead}}^r(\omega)}
\label{eq:recursion_relation}
\end{equation}

Selecting the physically meaningful solution (decaying for $\omega \to \infty$), we obtain:
\begin{equation}
g^r_{\text{lead}}(\omega)=\frac{1}{2\bar{\gamma}^{2}}\left(\omega-i\sqrt{4\bar{\gamma}^{2}-\omega^{2}}\right).
\end{equation}

\subsection{Effective 3-Site model and retarded Green's function}

We now compute the non-interacting retarded green function of the Anderson model $g^r$. Integrating out the electrode degrees of freedom reduces the system to three sites: the central impurity (site 0) and adjacent lead sites (-1 and 1). The retarded Green's function is therefore a matrix $g^r(\omega) \in \mathbb{C}^{3\times3}$, satisfying Dyson equation:

\begin{equation}
g^r(\omega)=\frac{1}{\omega-\epsilon_{d}-\Delta(\omega)}
\label{eq:retarded_green}
\end{equation} 

where the hybridization matrix $\Delta(\omega)$ couples the impurity to the leads:

\begin{equation}
\Delta(\omega)  =
\left(
\begin{array} {ccc }0 & \gamma g^r_{\text{lead}}(\omega) & 0\\
\gamma g^r_{\text{lead}}(\omega) & 0 & \gamma g^r_{\text{lead}}(\omega)\\
0 & \gamma g^r_{\text{lead}}(\omega) & 0
\end{array} \right)
\label{eq:hyb_func}
\end{equation}
Explicit matrix inversion of Eq.~\eqref{eq:retarded_green} using Eq.~\eqref{eq:recursion_relation} yields:

\onecolumngrid
\noindent\rule{10cm}{0.4pt}
\begin{equation}
g^{r}(\omega)  =\frac{1}{\omega-E_{d}-2\gamma^2 g_{\text{lead}}^r(\omega)}\times
\begin{pmatrix} g_{\text{lead}}^r(\omega)\left[\omega-E_{d}-\gamma{{}^2}g_{\text{lead}}^r(\omega)\right] & \gamma g_{\text{lead}}^r(\omega) & \gamma^{2}\left[g_{\text{lead}}^r(\omega) \right]^2\\
\gamma g_{\text{lead}}^r(\omega) & 1 & \gamma g_{\text{lead}}^r(\omega)\\
\gamma^{2}\left[g_{\text{lead}}^r(\omega) \right]^2 & \gamma g_{\text{lead}}^r(\omega) &g_{\text{lead}}^r(\omega)\left[\omega-E_{d}-\gamma^2 g_{\text{lead}}^r(\omega)\right]
\end{pmatrix}
\label{eq:green_retarded_explicit}
\end{equation}
\twocolumngrid

\subsection{Lesser and greater Green's functions in energy}
From the retarded Green, one computes the lesser and greater Green functions $g^\lessgtr(\omega)$ containing the non-equilibrium physics. One introduces $f_{l/r}^{<}(\omega)$ the Fermi function on the left and right electrode and $f_{l/r}^{>}(\omega)=1-f_{l/r}^{<}(\omega)$ giving the statistics of holes. The link between the greater/lesser and retarded/advanced Green under the hypothesis that the electrodes stay in equilibrium (\cite{Waintal24} Eq.[199]) is given by:
\begin{equation}
g^\lessgtr(\omega)=g^r(\omega) \Sigma^\lessgtr(\omega) g^a(\omega)
\end{equation} 
where $g^a = [g^r]^\dagger$ and the self-energies are:

\begin{equation}
\Sigma^\lessgtr(\omega)  = \mp 2i\Im(g_{\text{lead}}^r(\omega))
\begin{pmatrix}f_l(\omega)^\lessgtr & 0 & 0\\
0 & 0 & 0\\
0 & 0 & f_r(\omega)^\lessgtr
\end{pmatrix}
\end{equation}
In order to facilitate calculations one introduces
\begin{align}
	\eta(\omega)&=-(\omega-E_{d}-\gamma{{}^2}g(w))\\
	\xi(\omega)&=\gamma^{2}g(\omega)
\end{align} 
then, using \eqref{eq:green_retarded_explicit} one gets: 
\onecolumngrid
\noindent\rule{10cm}{0.4pt}
\begin{equation}
\label{eq:green_energy}
g^{\lessgtr}(\omega)  = p^{\lessgtr} (\omega) 
\begin{pmatrix}\frac{f_{l}^{\lessgtr}(\omega)\left|\eta(\omega)\right|^{2}+f_{r}^{\lessgtr}(\omega)\left|\xi(\omega)\right|^{2}}{\gamma^{2}} & \frac{-f_{l}^{\lessgtr}(\omega)\eta(\omega)+f_{r}^{\lessgtr}(\omega)\xi(\omega)}{\gamma} & -\frac{f_{l}^{\lessgtr}(\omega)\xi(\omega)^{*}\eta(\omega)+f_{r}^{\lessgtr}(\omega)\xi(\omega)\eta^{*}(\omega)}{\gamma^{2}}\\
\frac{-f_{l}^{\lessgtr}(\omega)\eta^{*}(\omega)+f_{r}^{\lessgtr}(\omega)\xi^{*}(\omega)}{\gamma} & f_{l}^{\lessgtr}(\omega)+f_{r}^{\lessgtr}(\omega) & \frac{f_{l}^{\lessgtr}(\omega)\xi^{*}(\omega)-f_{r}^{\lessgtr}(\omega)\eta^{*}(\omega)}{\gamma}\\
-\frac{f_{l}^{\lessgtr}(\omega)\xi(\omega)\eta^{*}(\omega)+f_{r}^{\lessgtr}(\omega)\xi(\omega)^{*}\eta(\omega)}{\gamma^{2}} & \frac{f_{l}^{\lessgtr}(\omega)\xi(\omega)-f_{r}^{\lessgtr}(\omega)\eta(\omega)}{\gamma} & \frac{f_{l}^{\lessgtr}(\omega)\left|\xi(\omega)\right|^{2}+f_{r}^{\lessgtr}(\omega)\left|\eta(\omega)\right|^{2}}{\gamma^{2}}
\end{pmatrix}
\end{equation}
\twocolumngrid Where the prefactor $p^{\lessgtr}$ is defined by 
\begin{equation}
p^{\lessgtr} = \frac{\pm i\sqrt{4\bar{\gamma}^{2}-\omega^{2}}\theta(2\bar{\gamma}-|w|)\gamma^{2}}{(\bar{\gamma}^{2}-2\gamma^{2})\left(\omega-D_{1}\right)\left(\omega-D_{2}\right)}
\end{equation}
and its singularities $D_{1/2}$ are:
\begin{equation}
D_{1/2}=\frac{E_{d}(\bar{\gamma}^{2}-\gamma^{2})\mp\gamma^{2}\sqrt{E_{d}^{2}+8\gamma^{2}-4\bar{\gamma}^{2}}}{\bar{\gamma}^{2}-2\gamma^{2}}
\end{equation}

\subsection{Flat-Band approximation}

All the calculation will be performed in the zero temperature limit such that
\begin{equation}
f_{l/r}^{<}=\theta(\omega-\mu_{l/r})
\end{equation} 
with $\mu_{l/r}$ the chemical potential in the right or left electrode. 
Furthermore one adopts the flat-band limit: the coupling energy on the electodes $\bar{\gamma}$ is supposed to be much higher than the other energy scales of the system $\bar{\gamma}\gg\gamma,\epsilon_{d},\mu_{l},\mu_{r},\omega$ so that
\begin{equation}
\sqrt{4\bar{\gamma}^{2}-\omega^{2}}  \approx 
2\bar{\gamma}
\end{equation}
Therefore, the hybridization function \eqref{eq:hyb_func} does no longer depends on energy. We define the characteristic energy scale $\Gamma$, which naturally arises from the hybridization strength:
\begin{equation}
\Gamma \equiv \gamma g(\omega) \approx \frac{2\gamma^{2}}{\bar{\gamma}}
\end{equation}
In this limit one gets the following simplifications for $\xi$, $\eta$ $D_{1/2}$ and $p^\lessgtr$ as a function of $\Gamma$:

\begin{align}
	\xi(\omega) &=-i\frac{\Gamma}{2} \\
	\eta(\omega)&=E_{d}-\omega-i\frac{\Gamma}{2}\\
	D_{1/2}&= E_d \pm i \Gamma \\
	p^\lessgtr(\omega)&=\frac{\pm  i \Gamma}{(\omega-E_d-i\Gamma)(\omega-E_d+i\Gamma)}
\end{align}

\subsection{Flat-band lesser and greater Green's functions in time}

In order to compute the non-interacting
green function in time, one needs to perform the Fourier transform :
\begin{equation}
g^{\lessgtr}(t)=\frac{1}{2\pi}\int_{-\infty}^{+\infty}g^{\lessgtr}(\omega)e^{-i\omega t}d\omega
\label{eq:fourier_transform_green_time}
\end{equation}
To compute the charge and the currents, we need the components of the green matrix where at least one index is the central site $0$ (e.g. $g^\lessgtr_{-10}$, $g^\lessgtr_{01}$, ect. see \cite{Profumo15}). From \eqref{eq:green_energy}, each of this components contains terms of the following form :
\begin{align}
	p^\lessgtr(\omega) &=\pm \left[ \frac{1}{2}\frac{1}{\omega-E_d-i\Gamma}-\frac{1}{2}\frac{1}{\omega-E_d+i\Gamma} \right]\\
	p^\lessgtr(\omega)\xi(\omega) &=\mp \left[ \frac{i\Gamma}{4}\frac{1}{\omega-E_d-i\Gamma}-\frac{i\Gamma}{4}\frac{1}{\omega-E_d+i\Gamma} \right]\\
	p^\lessgtr(\omega)\xi(\omega)^* &=\pm \left[ \frac{i\Gamma}{4}\frac{1}{\omega-E_d-i\Gamma}-\frac{i\Gamma}{4}\frac{1}{\omega-E_d+i\Gamma} \right]\\			
	p^\lessgtr(\omega)\eta(\omega) &=\mp \left[ \frac{3i\Gamma}{4}\frac{1}{\omega-E_d-i\Gamma}+\frac{i\Gamma}{4}\frac{1}{\omega-E_d+i\Gamma} \right]\\
	p^\lessgtr(\omega)\eta(\omega)^* &=\mp \left[ \frac{i\Gamma}{4}\frac{1}{\omega-E_d-i\Gamma}+\frac{3i\Gamma}{4}\frac{1}{\omega-E_d+i\Gamma} \right]
\end{align}

Consequently, Green functions in time can be expressed in terms of the functions $I^{\lessgtr}(t,\mu,D)$, where $D$ is a complex number.
\begin{align}
I^{<}(t,\mu,D) & =\int_{-\infty}^{\mu}\frac{1}{\omega-D}e^{-i\omega t}d\omega\\
I^{>}(t,\mu,D) & =-\int_{\mu}^{\infty}\frac{1}{\omega-D}e^{-i\omega t}d\omega
\end{align}
An analytical formula can be obtained for $I^{\lessgtr}(t,\mu,D)$. As an example, one derives $I^<$ while the same can be applied to $I^>$. There are three cases to examine, depending on the value of $t$.

\paragraph{Case $t>0$}

\begin{figure}
	
	\centerline{
		\includegraphics[scale=0.5]{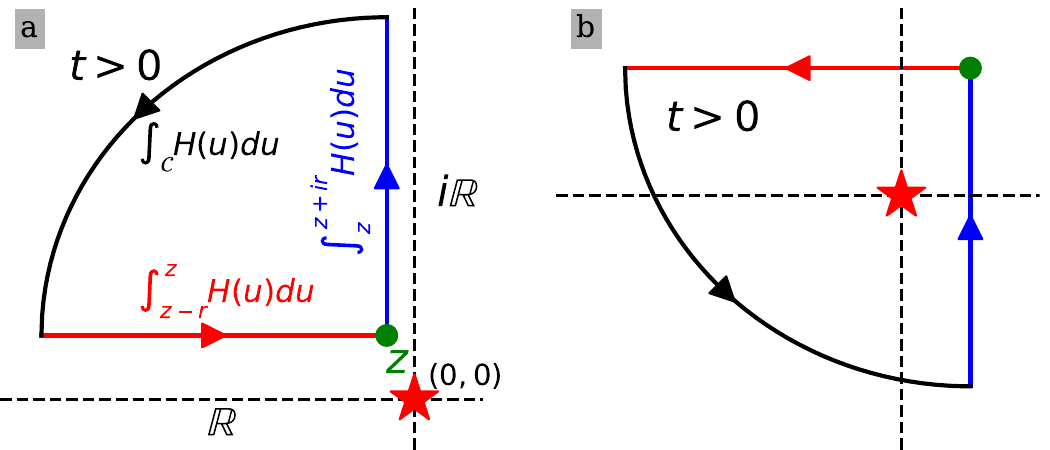}
	}
	\caption{Contour integration of equation \eqref{eq:contour_integ}. \textit{a)} case of $t>0$, the point $z$ of integration is in green, the singularity of $H$ is the red star. The contour is decomposed in three curves: the integral of interest (blue), the vanishing contour (black) and the $E_1$ related part (red). \textit{b)} same consideration for $t<0$. Note that the presence of a residus in the equation \eqref{eq:contour_integ} depend on the position of $z$ with respect to the origin, e.g. in (a) we illustrate the situation with the singularity outside the contour, it is also possible to have it inside as in (b).
		\label{fig:contour_integ}}
\end{figure}

 Let's consider first $t>0$. Using the change of variable $u=-it(\omega-D)$ one gets:

\begin{equation}
I^{<}(D,\mu,t)=-e^{-itD}\int^{+i\infty}_{it(D-\mu)}du\frac{e^{u}}{u}
\end{equation}
where the integral is performed along the parallel to the imaginary axis. For clarity, we define $z=it(D-\mu)$. One can notice that the function
\begin{equation}
H(u)=\frac{e^{u}}{u}
\end{equation}
is analytical and exhibits one pole on the complex plain in $u=0$. By performing contour integration on a quarter of circle $C$ of radius $r$, (see figure \ref{fig:contour_integ}) one can apply the theorem of residue:
\begin{align}
	\label{eq:contour_integ}
	\int_{z}^{z+ir}\frac{e^{u}du}{u}&+\int_{C}\frac{e^{u}du}{u}+\int_{z-r}^{z}\frac{e^{u}du}{u}\\&=2i\pi\theta(\Re(z))\theta(-\Im(z))\nonumber
\end{align}
As $r\to\infty,$ the integral on $C $ vanish. Indeed it is bounded by:
\begin{equation} |\int_{C}\frac{e^{u}du}{u}|\leq\int_{C}|\frac{e^{u}du}{u}|\leq\frac{\left|e^{z}\right|}{r}\int_{\theta=0}^{\pi/2}e^{-r\sin(\theta)}du
\end{equation}
using the inequality $\sin(\theta)>2\theta/\pi$ for $\theta \in [0,\pi/2]$, one gets:
\begin{equation} \int_{\theta=0}^{\pi/2}e^{-r\sin(\theta)}du \leq \int_{\theta=0}^{\pi/2}e^{-2r\theta/\pi}du\leq \frac{\pi}{2r}\underset{r\rightarrow \infty}{\rightarrow}0
\end{equation}
Therefore as $r\rightarrow\infty$, one can write:
\begin{equation}
\int_{z}^{+i\infty}\frac{e^{u}du}{u}=E_{1}(-z)+2i\pi\theta(\Re(z))\theta(-\Im(z))
\end{equation}
where the $E_1$ function can be evaluated efficiently \cite{Pegoraro11}, and is defined by:
\begin{equation}
E_1(z) = \int_z^\infty \frac{e^{-u}}{u}du=-\int^{-z}_{-\infty} \frac{e^{u}}{u}du
\end{equation}
so that for $t>0$ one has:
\begin{align} 
	I^<(D,\mu,t)=-e^{-itD}\left(E_{1}(-it(D-\mu)) \right.\\
	\left. +2i\pi\theta(-\Re(D-\mu))\theta(-\Im(D))\right) \nonumber
\end{align}

\paragraph{Case $t<0$}
For the case $t<0$, one consider an contour with a negative real part (see Fig \ref{fig:contour_integ}) one gets in a similar way:
\begin{equation}
\int_{z}^{-i\infty}\frac{e^{u}}{u}du=E_{1}(-z)-2i\pi\theta(\Re(z))\theta(\Im(z))
\end{equation}
such that :
\begin{align}
	I^{<}(D,\mu,t)=-e^{-itD}\left(E_{1}(-it(D-\mu))\right.\\
	\left.-2i\pi\theta(-\Re(D-\mu))\theta(\Im(D))\right) \nonumber
\end{align}

\paragraph{Case $t=0$} 
For $t=0$,
\begin{equation}
I^{<}(D,\mu,0)=\int_{-\infty}^{\mu}\frac{1}{\omega-D}d\omega
\end{equation}
By writting $-\omega-D=\rho e^{i\theta}$, one need to compute the limit of $ \log(-\omega-D)$ as $\omega\to\infty$. Remark that $\lim_{\omega\to\infty}\theta=\textrm{\ensuremath{-\pi}sgn}(\Im(D))$, therefore:

\begin{equation}
I^{<}(D,\mu,0)=\log(-D+\mu) - \lim_{\omega\to\infty}\log(\omega)+i\pi \text{sgn}(\Im D)
\end{equation}

The divergent term \(\lim_{\omega\to\infty} \log(\omega)\) cancels when considering \(g^{\lessgtr}_{0,0}\) (the dot-dot component), as the divergence from \(I^<(D_1)\) is offset by \(I^<(D_2)\). However, the dot-lead components of the Green functions diverge at \(t = 0\) due to the flat-band limit: at \(t = 0\), the dot interacts instantaneously with all lead states, causing an unphysical infinite response. Despite this, the physical current---defined by the real part of the dot-lead Green function~\cite{Haldane78}---remains finite at every order in Keldysh perturbation. Thus, the \(\lim_{\omega\to\infty} \log(\omega)\) term is safely discarded.

The factor \(\frac{1}{\gamma}\) in the dot-lead Green functions becomes infinitesimal as \(\gamma \to 0\). However, since it is counterbalanced by a \(\gamma\)-dependent term in the current formula (Eq.~\eqref{eq:green_current}), it can be set to 1. (A rigorous derivation using Wick's theorem for each order confirms this.)

As a result, the non-interacting Green function in the Flat-Band limit is given by:

\onecolumngrid
\noindent\rule{9cm}{0.4pt}
\begin{align}
	I^<(D,\mu,t)&=
	\begin{cases}
		-e^{-itD}\left(E_{1}(-it(D-\mu))+2i\pi\textrm{sgn}(t)\theta(-\Re(D)+\mu)\theta(-\Im(tD))\right)  &\textrm{if }t\neq0\\
		i\pi sgn(\Im D)+\log(-D+\mu) &\textrm{if } t=0
	\end{cases} \\
	I^{>}(D,\mu,t) &=
	\begin{cases}
		-e^{-itD}\left(E_{1}(-it(D-\mu))-2i\pi\textrm{sgn}(t)\theta(-\Re(D)+\mu)\theta(-\Im(tD))\right) & \textrm{if }t\neq0\\
		\log(-D+\mu) &\textrm{if } t=0
	\end{cases}
\end{align}
\begin{align}
	g^{\lessgtr}_{0,0}&= \frac{1}{2\pi}\left(\pm \left[\frac{1}{2}I^<(t,\mu_l,E_d+i\Gamma) - \frac{1}{2}I^<(t,\mu_l,E_d-i\Gamma)\right] \pm \left[\frac{1}{2}I^<(t,\mu_r,E_d+i\Gamma) - \frac{1}{2}I^<(t,\mu_r,E_d-i\Gamma)
	\right]\right)\\
	g^{\lessgtr}_{-1,0} &= \frac{1}{2\pi}\left(\pm \left[ \frac{3i\Gamma}{4}I^<(t,\mu_l,E_d+i\Gamma) + \frac{i\Gamma}{4}I^<(t,\mu_l,E_d-i\Gamma)\right] \mp \left[ \frac{i\Gamma}{4}I^<(t,\mu_r,E_d+i\Gamma) - \frac{i\Gamma}{4}I^<(t,\mu_r,E_d-i\Gamma)\right]\right) \\
	g^{\lessgtr}_{0,-1} &=\frac{1}{2\pi}\left( \pm \left[ \frac{i\Gamma}{4}I^<(t,\mu_l,E_d+i\Gamma) + \frac{3i\Gamma}{4}I^<(t,\mu_l,E_d-i\Gamma)\right] \pm \left[ \frac{i\Gamma}{4}I^<(t,\mu_r,E_d+i\Gamma) - \frac{i\Gamma}{4}I^<(t,\mu_r,E_d-i\Gamma)\right]\right)\\
	g^{\lessgtr}_{-1,0} &=\frac{1}{2\pi}\left( \pm \left[ \frac{3i\Gamma}{4}I^<(t,\mu_r,E_d+i\Gamma) + \frac{i\Gamma}{4}I^<(t,\mu_r,E_d-i\Gamma)\right] \mp \left[ \frac{i\Gamma}{4}I^<(t,\mu_l,E_d+i\Gamma) - \frac{i\Gamma}{4}I^<(t,\mu_l,E_d-i\Gamma)\right]\right) \\
	g^{\lessgtr}_{0,-1} &= \frac{1}{2\pi}\left(\pm \left[ \frac{i\Gamma}{4}I^<(t,\mu_r,E_d+i\Gamma) + \frac{3i\Gamma}{4}I^<(t,\mu_r,E_d-i\Gamma)\right] \pm \left[ \frac{i\Gamma}{4}I^<(t,\mu_l,E_d+i\Gamma) - \frac{i\Gamma}{4}I^<(t,\mu_l,E_d-i\Gamma)\right]\right)
\end{align}
\twocolumngrid
\section{Validity of the Flat-Band approximation}

This appendix verifies the calculations of Section~\ref{sec:Non-interacting Green} and evaluates the validity of the flat-band approximation. For finite bandwidths, the lesser/greater Green's functions $g^\lessgtr(t)$ lack closed-form expressions and require numerical evaluation of the oscillatory integral in Eq.~\eqref{eq:fourier_transform_green_time} from Eq.~\eqref{eq:green_energy}. The integrand's rapid oscillations (Fig.~\ref{fig:comp_flat}a) render conventional quadrature methods ineffective. We present two robust approaches for computing this integral: (1) contour integration via the residue theorem and (2) tensor cross-interpolation (TCI) in quantics mode. Using these methods, we quantify the accuracy of the flat-band limit as a function of the bandwidth parameter $\Gamma$.
All calculations use the following parameters:
\begin{itemize}
	\item $\gamma = 0.01$, $\bar{\gamma} = 1$, $\Gamma = 2\gamma^2/\bar{\gamma} = 2 \times 10^{-4}$
	\item $\epsilon_d = -4\Gamma$, $\mu_l = 2\Gamma$, $\mu_r = -3\Gamma$
	\item Times are expressed in units of $1/\Gamma$
	\item Dot-lead Green's functions normalized by $\Gamma/\gamma$ for comparison with flat-band results
\end{itemize}

\subsection*{Method 1: Contour integration via residue theorem}
The components of $g^\lessgtr(t)$ can be expressed as sums of integrals:
\begin{equation}
I[\zeta,a,b]=\frac{1}{2\pi}\int_{a}^{b}\frac{\pm i\sqrt{4\bar{\gamma}^{2}-\omega^{2}}\gamma^{2}\zeta(\omega)e^{-i\omega t}d\omega}{(\bar{\gamma}^{2}-2\gamma^{2})\left(\omega-D_{1}\right)\left(\omega-D_{2}\right)}
\label{eq:contour}
\end{equation}
where $\zeta \in \left\{ \xi,\xi^*,\eta,\eta^*,1 \right\}$.  $a, b$ are determined by band edges ($\omega \in [-2\bar{\gamma}, 2\bar{\gamma}]$) and occupation functions $f^\lessgtr(\omega)$. These integrals are evaluated using the residue theorem in the complex plane (see Fig. \ref{fig:comp_flat}b). When $t$ is positive (negative), integration is performed in the lower (upper) complex half-plane along an infinite contour. The two vertical branches are calculated through a change of variable $u=\frac{\omega}{1+\omega}$ in order to keep finite integration bounds and using Chebysehv quadrature. Depending on the half-plane considered and the integration limits, there may or may not be a residual to take into account at $D_1$ or $D_2$ .

\subsection*{Method 2: Tensor cross-interpolation (TCI) in quantics mode}
We employ the Tensor Cross-Interpolation algorithm in quantics mode (via the \texttt{xfacpy} module~\cite{NunezFernandez24}) to approximate \( g^\lessgtr(\omega)e^{-i\omega t} \) on a \( 2^{30} \)-point grid. The subsequent integration is then performed directly using the tensor train structure. Across all time values, this interpolation maintains a low rank (\(\chi \sim 10\)). We define the normalized average error \( E(\chi) \) over \( t \in [-10,10] \) as:
\begin{equation}
E(\chi) = \frac{\int_{-10}^{10} \left|g^<_{0,1}(t) - g_\chi(t)\right| \, dt}{\int_{-10}^{10} \left|g^<_{0,1}(t)\right| \, dt},
\end{equation}
where \( g_\chi(t) \) denotes the rank-\(\chi\) tensor train approximation of \( g^<_{0,1}(t) \). As shown in Fig.~\ref{fig:comp_flat}(c), this error decreases rapidly with increasing \(\chi\).

\subsection*{Validity of the flat-band limit}
\begin{figure*}
	
	\centerline{
		\includegraphics[scale=0.60]{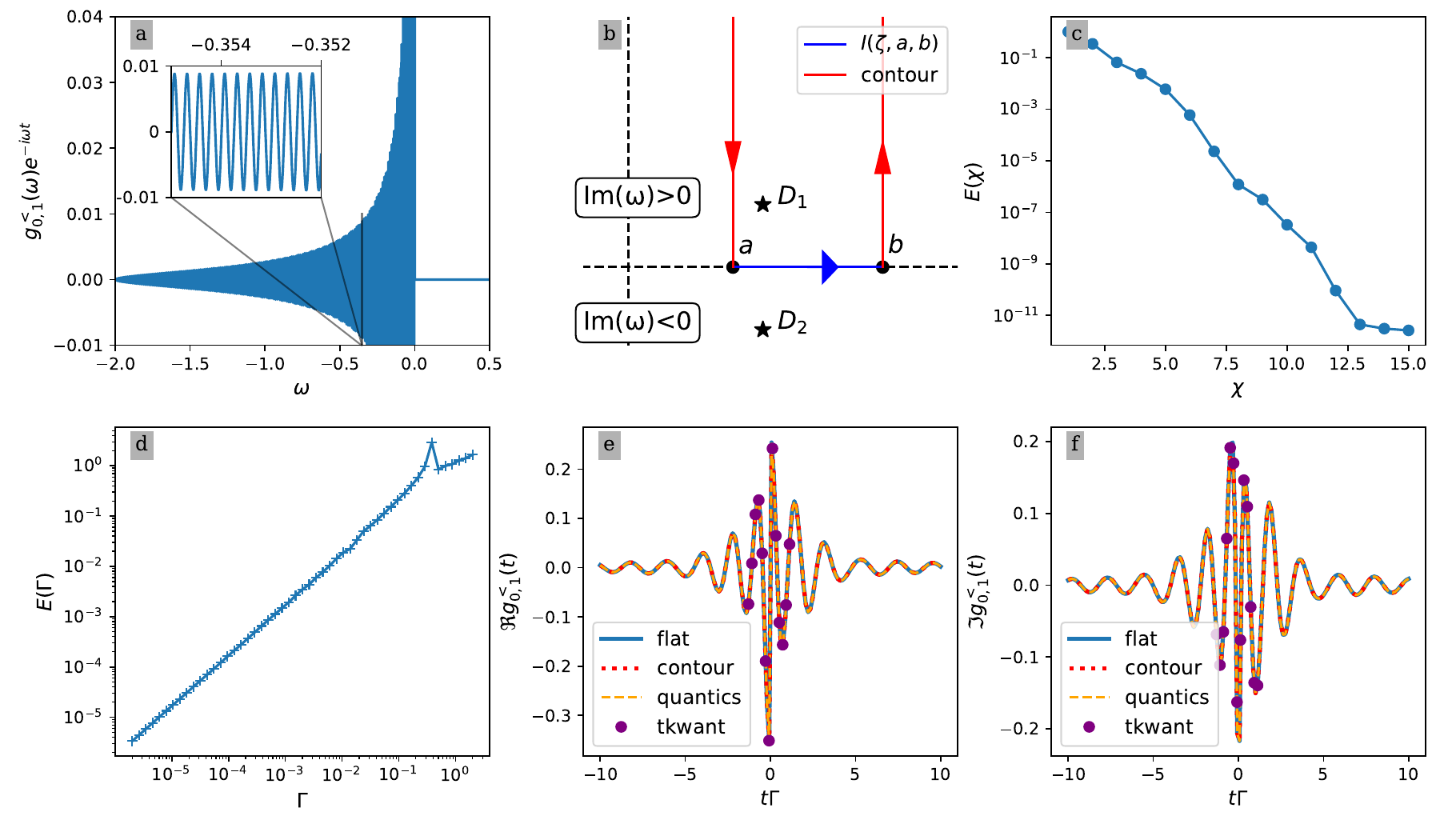}
	}
	\caption{Assessing the validity of the Flat-Band limit approximation. \textit{a)} $g^<_{0,1}(\omega)e^{-i\omega t}$ at $t=4$ as a function of $\omega$ with $t = 4$. \textit{b)} Complex contour used to evaluate the integral in Eq. \eqref{eq:contour}. \textit{c)} Numerical error $E(\chi)$ as a function of $\chi$ when computing $g^<_{0,1}(t=4)$ via quantics integration. \textit{d)} Error $E(\Gamma)$ as a function of $\Gamma$, testing the robustness of the approximation. \textit{e)} and \textit{f)} real and imaginary part of $g^<_{0,1}$ as a function of time, comparing results from analytical, flat-band, quantics and tkwant integration methods. Panels d–f collectively benchmark the accuracy of the flat-band limit.}
	\label{fig:comp_flat}
\end{figure*}

We now vary $\Gamma$ with $g^<_{0,1}(t)$ obtained by one of the above methods. We define the deviation with respect to the flat-band by:

\begin{equation}
E(\Gamma) = \frac{\int_{t=-10}^{10}  \left|g^<_{0,1}(t,\Gamma=0)-g^<_{0,1}(t,\Gamma)\right|dt}{\int_{t=-10}^{10}\left|g^<_{0,1}(t,\Gamma=0)\right|dt}
\end{equation}
where $g^<_{0,1}(t,\Gamma=0)$ is the flat-band limit and $g^<_{0,1}(t,\Gamma)$ is the model with a finite value of $\Gamma$. From figure \ref{fig:comp_flat}\textit{d}) we check that the flat-band approximation is a first-order in the sense that this deviation varies linearly with $\Gamma$.

Ultimately, we compare $g^<_{0,1}(t)$ computed using the flat-band limit, contour integration, quantics integration, and numerical results from \texttt{tkwant} \cite{Kloss21}, presented in panels (e) (real part) and (f) (imaginary part).

\end{document}